\documentclass[aps,pra,superscriptaddress,notitlepage,a4paper,nofootinbib]{revtex4-2}
\usepackage[utf8]{inputenc}
\usepackage{tikz}
\usepackage{amsmath,amssymb,amsfonts,graphics,graphicx,dcolumn,bm,enumerate}
\usepackage[pagewise]{lineno}  
\usepackage{ftnxtra}
\usepackage{fnpos}
\usepackage{comment,natbib,appendix}
\usepackage{multirow,color}
\usepackage{chngpage}
\usepackage{afterpage}
\usepackage{xcolor}
\usepackage{amsthm}
\usepackage{natbib}
\usepackage{hyperref}
\usepackage[margin=1in]{geometry}
\usepackage{epstopdf}
\usepackage{float}
\usepackage{soul}
\usepackage{caption}
\usepackage{subcaption}
\newcommand{\snb}
{\affiliation{S. N. Bose National Centre for Basic Science, Kolkata 700106, India.}}

\newcommand{\isi}
{\affiliation{Indian Statistical Institute, Kolkata 700108, India.}}

\begin{document}

\title{Exploring Citation Diversity in Scholarly Literature: An Entropy-Based Approach}

\author{Suchismita Banerjee}
\email[Email: ]{suchib.1993@gmail.com}
\snb \isi

\author{Abhik Ghosh}
\email[Email: ]{abhik.ghosh@isical.ac.in}
\isi

\author{Banasri Basu}
\email[Email: ]{sribbasu1@gmail.com}
\isi

\begin{abstract} 
This study explores the  citation diversity in scholarly literature, analyzing different patterns of citations observed within different countries and academic disciplines. We examine citation distributions across top institutions within certain countries and find that the higher end of the distribution follows a Power Law or Pareto Law pattern; the scaling exponent of the Pareto Law varies depending on the number of top institutions included in the analysis. By adopting a novel entropy-based diversity measure, our findings reveal that countries with both small and large economies tend to cluster similarly in terms of citation diversity. The composition of countries within each group changes as the number of top institutions considered in the analysis varies.
Moreover, we analyze citation diversity among award-winning scientists across six scientific disciplines, finding significant variations. 
We also explore the evolution of citation diversity over the past century across multiple fields. A
gender-based study in several disciplines confirms varying citation diversities among male and female scientists. 
Our innovative citation diversity measure stands out as a valuable tool for assessing  the unevenness of citation distributions, providing deeper insights that go beyond what traditional citation counts alone can reveal. This  comprehensive  analysis enhances  our understanding of global scientific contributions and fosters a more equitable view of academic achievements.
\end{abstract}

\maketitle


\section{Introduction}
Citations are the currency of academia, reflecting impacts and influences of research publications. Ideally, a fair citation landscape would see recognition distributed proportionally to the quality and contribution of research. 
The term `quality' generally refers to the rigor, originality, and reliability of the research, while `contribution of research' refers to the significance or impact the research has within and beyond academia on advancing knowledge, solving problems, or influencing a field of study. Both factors play a crucial role in determining why certain papers receive high citation counts. Nevertheless, citations are also influenced by external factors such as visibility, collaboration networks, and research trends, rather than just intrinsic quality and contribution.

However, in practice, citations are often unevenly distributed, with a small number of papers receiving a disproportionately large share of citations, while the majority receive far fewer– a phenomenon widely discussed as citation inequality~\citep{Garfield_1972,crespo_2013,nielsen_2021,dong_2021}.
Measuring the inequality in citation patterns has been a central focus of bibliometric studies, with many approaches borrowing from economic inequality metrics~\citep{gintropy,banerjee_2023}.
There are numerous studies~\citep{gintropy1,suchi_k-index,Leyd_2002,entropy_1980,Rajaram_2017,Nsakanda_2007,Jost_2006} on citation inequality using various inequality indices originally developed in economics (like Gini index), as well as entropy-based measures (like Shannon entropy)~\citep{Li_2015,Mug_2016}. 
Moreover, indices like the Hirsch index (h-index) have been utilized to summarize citation distributions~\citep{Lin_2020,h-index,h-index1,h-index2}.

$Citation~diversity$, on the other hand, measures the variety and evenness of the distribution of citations across different categories, which may be institutes, authors, disciplines, etc.~\citep{work_diversity,work1,work4}. Unlike inequality, which focuses on how citations are numerically distributed, diversity captures the breadth of influence a paper or institution has across multiple fields. High citation diversity indicates a wide-ranging impact over various areas, while low diversity suggests influence being concentrated within a narrow domain. 
It may be mentioned here that assessing diversity within a population is a crucial issue across various applied sciences, such as ecology, biology, economics, sociology, physics, and management sciences; see, e.g., \citep{work2,work3,Lein_2012,stirling_2007,Leyd_2019}. 
However, the potential of generalized entropy measures, which offer a versatile framework for assessing diversity, has not been fully explored in the context of citation diversity.
This gap presents an opportunity to use novel entropy measures for examining the breadth and evenness of citation distribuhavetions. 

This study introduces a two-parameter generalization of the Renyi entropy to measure citation diversity, emphasizing the evenness of citations across different categories rather than focusing on citation counts alone. The methodology builds on the concept of logarithmic norm entropy~\citep{Ghosh_2023,7}, adapted from information theory to quantify how evenly citations are spread within research domains, institutions, or disciplines. Higher entropy reflects a broader and more uniform citation distribution, while lower entropy indicates concentration within fewer categories. We apply this framework to explore variations in citation diversity globally, beginning with top-ranked academic institutions across different countries to explore potential geographical variations.
Then, we extend our investigation to analyze the diversity of citations received by publications of top award winning scientists (Nobel prize winners, Abel winners and Turing award winners). This analysis will encompass various disciplines, ensuring a holistic understanding of citation patterns in research publications across different 
academic fields. 
We also explore the time evolution of the citation diversity of the award winning scientists by analyzing the diversity of recent and century old award winners across various disciplines.
Finally, we dis-aggregate our findings by gender, enabling a nuanced exploration of potential gender-based disparities in citation practices of various academic disciplines. This entropy-based multifaceted approach on citation diversity offers a detailed picture, havecapturing the subtlety and variations in citation distribution within the scientific landscape.

This research facilitating an understanding of $citation$ $diversity$ in scholarly literature is presented in a clear and logical structure. 
In Section II, we outline data sources for citation information and meticulously describe the data employed in the study. We 
provide details regarding the selection criteria for the award winning scientists, the specific disciplines included, and the identification process for top institutes across different countries. 
Section III serves as the foundation of our analysis, providing a step-by-step analytical framework in  developing  the concept of general class of logarithmic norm entropy and its use as a measure of citation diversity.
Section IV presents the findings of our investigation where we delve into the analysis of citation diversity across various scenarios and the significance of our findings is also presented therein. 
Finally a concise summary of our key findings are provided in Section V, describing the main takeaways from our analyses. Furthermore, we have also discussed the broader implications of our research and potential avenues for future inquiry. 

\section{Data Description}
The {\it `Ranking Web of Universities}' (also commonly known as the \textit{Webometrics}) \citep{website1} is a comprehensive academic ranking system established in 2004, which appears twice per year since 2006. This public resource, developed by the Cybermetrics lab, encompasses over 31,000 higher education institutions or universities (referred to as the HEIs) across more than 200 countries. Webometrics employs a mix of webometric (all missions) and bibliometric (research mission) indicators to assess university performance, promoting open access to scholarly knowledge. 
It provides the detailed citation data of the top HEIs across the world through `Transparent Ranking: Top Universities by Citations in Top Google Scholar profiles' \citep{website}. We used the January 2024 edition  of this data (retrieved on April 1, 2024) for our citation analysis. The detailed data on citation counts of top institutes can be found in \citep{website1}.   

Moreover, our study also leverages data from {\it `Scopus'} \citep{scopus}, an extensive bibliographic database of peer-reviewed literature. 
We have used this resource for obtaining publication and citation information for award-winning authors across various disciplines.
This unified data source allows for robust comparisons and minimizes potential biases arising from using disparate data sources. To ensure consistency, 
we obtain total citation data for 21 scientists in each discipline from {\it `Scopus'} \citep{scopus} on May 23, 2024. Additionally, we collect publication and citation data for individual scientists, including 30 Nobel laureates in physics, chemistry, and physiology/medicine (split evenly between recent and century old awardees), 15 recent Abel prize winners in mathematics, 15 recent Turing award winners in computer science, and 15 recent Nobel laureates in economics from the same source~\citep{scopus} on May 23, 2024.

\section{Methodology}
\subsection{The General Class of logarithmic norm entropy (LNE) and diversity measure (D)}
It was long known that potential families of entropy measures can  be used as generalized diversity measures \citep{Rao}. Recently, the concept of logarithmic norm entropy (LNE) has been introduced in \citep{LNE_2018} as a new measure for quantifying diversity, justified by its better statistical efficiency and robustness properties compared to other existing classes of entropy based diversity measures. Building upon the established concept of Shannon entropy \citep{Shannon} and Renyi entropy \citep{Renyi}, the LNE offers a scale-invariant generalization of the latter \citep{7}. In this study, we leverage LNE to quantify citation diversity in scholarly literature. This unique approach allows us to assess the robustness of our findings and gain a more complete understanding of citation diversity patterns.

Consider a finite set of $M$ categories (denoted as $c_{1}, ..., c_{M}$) representing different domains of applications, such as citation patterns across top institutes, the author’s own publications, gender-based differences, and discipline-wise variations. The probability distribution over these categories is represented as $p=(p_{1},...,p_{M})$, where $p_{i}$ signifies the probability associated with category $c_{i}$ for each $i=1,...,M$. These probabilities are normalized to have the sum equal to one.
The value of $M$, the number of categories, depends on the specific context of the analysis.
The general classes of Shannon entropy and Renyi entropy are defined~\cite{7}, respectively, as:
\begin{equation}\label{eq1}
    H_{\beta}^{(S)}(p)= - \frac{\sum_{i}^{M}p_{i}^{\beta}\log p_{i}}{\sum_{i}p_{i}^\beta},~~~\text{where}~~\beta \in \mathbb{R}^+,
\end{equation}
\begin{equation}\label{eq2}
    H_{(\alpha,\beta)}^{(R)}(p) = \frac{1}{1 - \alpha}\log{\left[\frac{(\sum_{i}p_{i}^{\alpha + \beta -1})}{(\sum_{i}p_{i}^\beta)}\right]},~~~\text{where}~~\alpha,\beta \in \mathbb{R}^+
\end{equation}
Clearly, at $\beta = 1$, Eq.~(\ref{eq1}) and Eq.~(\ref{eq2}) coincides with the classical Shannon entropy and the Renyi entropy, respectively. 

Several other one and two-parameter generalizations of the entropy functional have been introduced in the literature, though their practical relevances and experimental validity vary. One notable example is a generalization of Renyi entropy, known as the Kapur’s generalized entropy~\cite{Kapur_1962,Kapur_1967} of order $\alpha$ and type $\beta$, which is defined as:
\begin{equation}\label{eq3}
    H_{(\alpha,\beta)}^{(K)}(p) = \frac{1}{\beta-\alpha}\log{\left[\frac{\sum_{i}p_{i}^{\alpha}}{\sum_{i}p_{i}^\beta}\right]},~~~\text{where}~~\alpha,\beta \in \mathbb{R}^+
\end{equation}
In this case also, when $\beta = 1$, Eq.~\ref{eq3} coincides with the Renyi entropy. This is a two-parameter generalization of Renyi entropy but it is important to note that neither the Renyi nor the Kapur's generalized entropy measures are scale-invariant.

In this study, we consider the novel scale-invariant generalization of the Renyi entropy, namely the LNE  measure defined as \citep{Ghosh_2023}
\begin{equation}\label{ln1}
 H_{(\alpha,\beta)}^{(LN)}(p) = \frac{\alpha \beta}{\beta - \alpha}\log{\left[\frac{(\sum_{i}p_{i}^\alpha)^{\frac{1}{\alpha}}}{(\sum_{i}p_{i}^\beta)^{\frac{1}{\beta}}}\right]},
\end{equation}
where $\alpha, \beta$ are two positive constants (tuning parameters) leading to different entropy measures. At $\beta = 1$ or $\alpha = 1$, the LNE reduces to the Renyi entropy family and is generally symmetric in the choice of $(\alpha, \beta)$.
We readily note the limiting interrelations between these entropies as:
\begin{equation*}
    \lim_{\alpha \rightarrow 1} H_{(\alpha,1)}^{(R)}(p) = \lim_{\alpha \rightarrow 1} H_{(\alpha,1)}^{(K)}(p) = \lim_{\alpha \rightarrow 1} H_{(\alpha,1)}^{(LN)}(p) = H_{1}^{(S)}(p) = - \sum_{i}p_{i}\log p_{i}.
\end{equation*}

Clearly, the maximum value of the diversity measure (D) equal $100\%$ for all members of the LNE family regardless of the values of the tuning parameters ($\alpha,\beta$). It always lies between $0$ to $100$ (both inclusive) with higher values indicating greater diversity and vice versa. The citation diversity measure (D), based on LNE, will be computed for each country and disciplinary group of prize winning scientists, as well as for individual award winners, replacing $p$ with its estimates $\widehat{p}$ derived from empirical data. We will also compute these metrics separately among males and females scientists within the award winning cohort.

Following (\ref{ln1}) and noting that the maximum possible value of all these entropies is $\log M$ for a model with $M$ categories, one can define the general Diversity measure (expressed in percentage for convenience) as:
\begin{equation}
 D= \frac{H_{(\alpha,\beta)}^{(LN)}(p)}{\log M} \times 100\%.   
\end{equation}

For computation of $D$ in different such domains of applications, $M$ takes different values. For example, when analyzing top institutes, $c_1, c_2, \ldots, c_M$ represents the set of institutes, and $M$ is the number of institutes considered within each country (i.e., M=10 while studying citation diversity among top 10 institutes, and so on) and the citation diversity, $D$, is then computed based on the citation data of these $M$ institutes for each country. The number of countries considered is dependent on the data availability and hence it varies depending on the ranking level. For top 10 institutes, we could include 72 countries, based on data availability, while for the study of top 20 institutes we could only use data from 55 countries, and similarly 25 countries for analyzing diversity among top 50 institutes.

\subsection{Asymptotic standard error 
and confidence interval}
Since we are estimating the LNE based diversity measures from empirical data, we must additionally quantify the extent of statistical errors associated with our estimates to draw more effective conclusions. As proved in \citep{Ghosh_2023}, such estimates of the diversity measure (D) will be $\sqrt{n}$-consistent and asymptotically normal with the asymptotic variance being $\frac{\sigma_{(\alpha,\beta)}^2(p)}{n(\log M)^2}$, where
\begin{equation*}
    \sigma_{(\alpha,\beta)}^{2}(p) = \frac{\alpha^2 \beta^2}{(\beta -\alpha)^2} \left[\frac{W_{2\alpha -1}}{W_{\alpha}^2} + \frac{W_{2\beta -1}}{W_{\beta}^2} - \frac{2W_{\alpha + \beta -1}}{W_{\alpha}W_{\beta}}\right],
\end{equation*}
with the notation $W_c(p) = \sum_{i=1}^{M}p_{i}^{c}$ for any $c > 0$. 
Note that, $\sigma_{(\alpha,\beta)}^{2}(p)$ is symmetric in the choice of $(\alpha, \beta)$, as intuitively expected from similar behavior of the LNE measure itself. 
Since $\sigma^2_{(\alpha,\beta)}(p)$ varies continuously in the citation distribution ($p$), we can reliably estimate it using our empirical data by replacing $p$ by its estimates $\widehat{p}$. Finally, taking square root of the estimated asymptotic variances, we get the (asymptotic) standard error (say $s$) of the estimated diversity measure (D), with lower values indicating more reliable diversity estimates.

By utilizing the standard errors (s) of the estimated diversity (D) in all our cases, we have computed and plotted the $95\%$ confidence intervals for the diversity measures as given by $(D-1.96s,D+1.96s)$. This formula is obtained from the standard theory of statistical inference by utilizing the asymptotic normality of the diversity estimate. Note that, the length of the confidence interval is directly proportional to the standard error, and hence indicates the reliability of the estimated diversity; shorter the confidence interval more reliable our estimates are. Moreover, such confidence intervals also help us to statistically compare the diversity measures for two contexts (e.g., countries, subjects, or scientists); two diversity values can be inferred to be significantly different at $5\%$ level if the associated $95\%$ confidence intervals do not overlap. This gives us a simple visual way to identify contexts having statistically similar or dissimilar diversities by just comparing the plots of their confidence intervals as presented in the following sections.
Throughout our entire analysis, we have used $\alpha = 2.0$ and $\beta = 0.5$ in the definition of the LNE based diversity. Although there exists a whole class of LNE measures with different choices of $\alpha$ and $\beta$ tailored to various applications, this specific combination is found to provide the most meaningful results for our citation data, along with having lower standard errors and narrower confidence intervals; it was also recommended in~\citep{Ghosh_2023} from statistical considerations.

\section{
Results and Discussions}
\subsection{Citation analysis among 
top institutes within different countries}

There is an increasing interest in global university rankings through various metrics \citep{Aguillo_2010}. The rich and exhaustive data of $university$ $rankings$ motivates one to analyse it from different  angles and various perspectives \citep{scientometric}. In the present investigation of $citation$ $diversity$ we consider the ranking of the universities/ institutes, ($N_I$), according to their total number of citations ($N_{cit}$). This helps us in examining the distribution pattern of total citations, across all disciplines, of top institutions of a country as well as the citation diversity measure (D) among the top institutions of each country. This study helps in delineating  the geographical variation \citep{Gomez_2022} of research activities. 

\subsubsection{Distribution pattern of citation counts of worldwide top institutes}

\begin{figure}[h]
     \centering
     \begin{subfigure}[b]{0.47\textwidth}
         \centering
         \includegraphics[width=\textwidth]{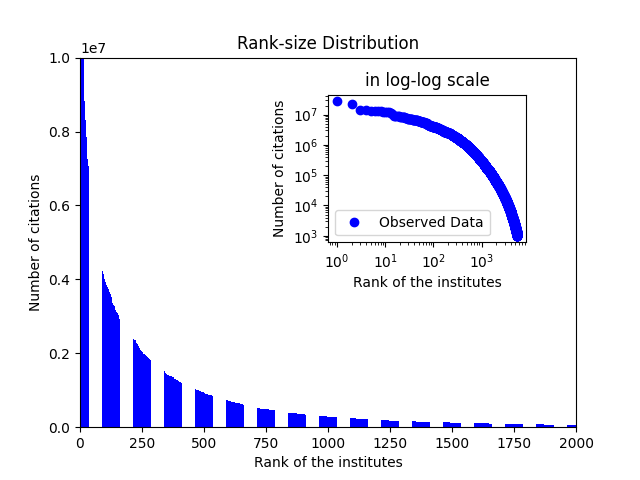}
         \caption{Bar plot of rank-size distribution, and its log-log plot in the inset.}
         \label{fig:dist1}
     \end{subfigure}
     \hfill
     \begin{subfigure}[b]{0.47\textwidth}
         \centering
         \includegraphics[width=\textwidth]{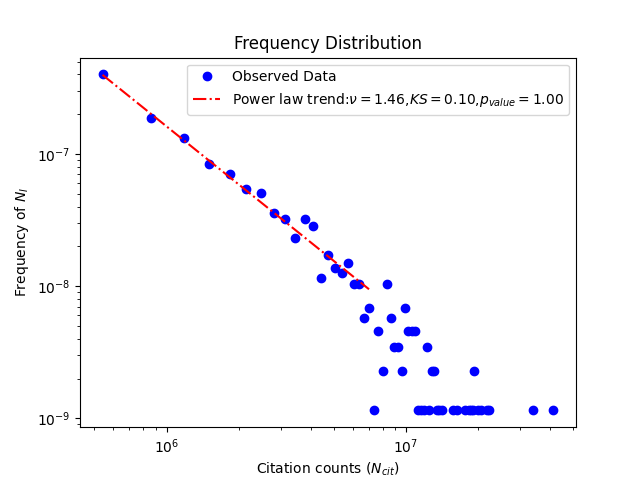}
         \caption{Frequency distribution and the fitted power law  in log-log scale}
         \label{fig:dist2}
     \end{subfigure}
        \caption{Rank-size and  frequency distribution plots of the total citation counts with the corresponding ranks of the institutes.
        The KS distance (KS),  $p$-value and the  exponent ($\nu$) of the fitted power law distribution is given in the inset of (b). }
        \label{fig:dist}
\end{figure}

Initially, we examine the distribution pattern of total citation counts $N_{cit}$ corresponding to a large number of  institutes/universities ($N_I$) from various countries around the world.  
According to the Webomterics data~\citep{website1}, considered for the analysis, rank 1 institute is the Harvard University of USA with citation count 27589889 and the Institute of Technology and Business of Czech Republic corresponds to rank 5661 with citation count 1004. The data furnishing a wide range of variation in $N_{cit}$. 
The distribution pattern (Fig.~\ref{fig:dist}) provides  valuable insights in understanding  how the citation data is spread out or clustered around the world's leading universities and institutes. 
Fig.~\ref{fig:dist1} is the bar plot for the rank-size distribution, based on  the ranks of the institutes $N_I$ and the corresponding size of citation counts $N_{cit}$; with an inset displaying the same plot in log-log scale for a clear understanding of the trend.  
Fig.~\ref{fig:dist2} depicts the frequency distribution curve of total citation counts across different institutions in log-log scale. The distribution plot exhibits a power-law behavior in the higher citation end. 
The robustness of the fitted power law is checked by a goodness-of-fit test yielding satisfactory Kolmogorov-Smirnov distance $(KS)$ and $p$-value for the fit. 

\begin{figure}[h]
     \centering
     \begin{subfigure}[b]{0.32\textwidth}
         \centering
    \includegraphics[width=\textwidth]{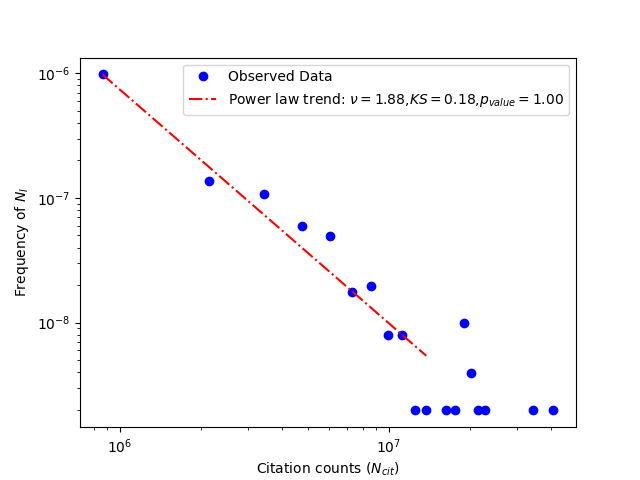}
         \caption{
         Top 10 institutes } 
         \label{fig:power1}
     \end{subfigure}
     \hfill
     \begin{subfigure}[b]{0.32\textwidth}
         \centering
         \includegraphics[width=\textwidth]{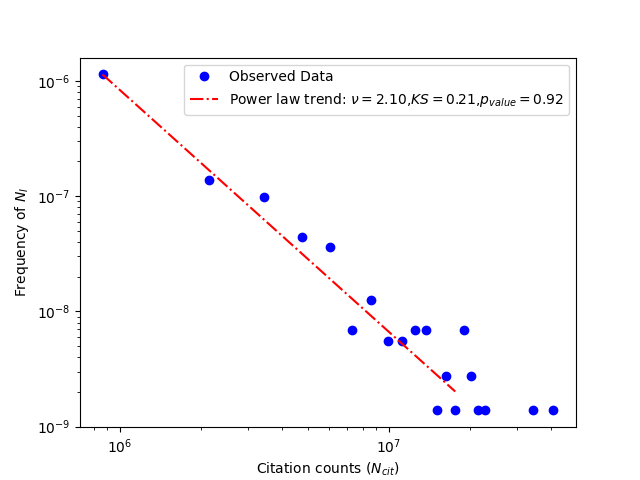}
         \caption{
         Top 20 institutes} 
         \label{fig:power2}
     \end{subfigure}
     \hfill
     \begin{subfigure}[b]{0.32\textwidth}
         \centering
         \includegraphics[width=\textwidth]{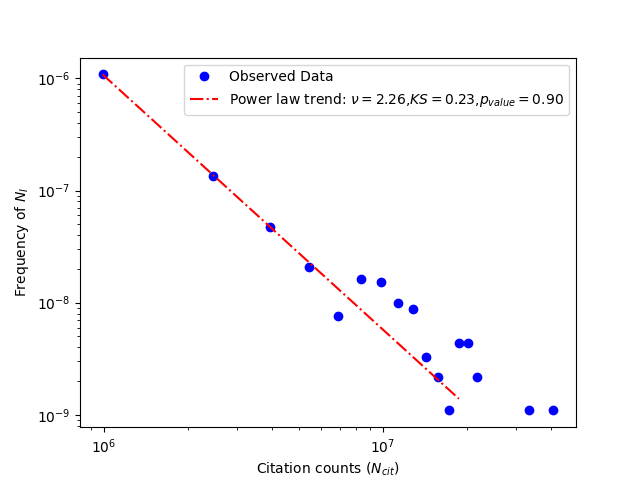}
         \caption{
         Top 50 institutes } 
         \label{fig:c}
     \end{subfigure}
        \caption{The 
        citation distribution for top 10, 20 and 50 institutes in log-log scale for various countries across the world. 
        Blue dots are the observed data points and red line represents the fitted power law with varying exponents ($\nu$), KS distance (KS) and the $p$-values in each sub-plot.}
        \label{fig:pow}
\end{figure}

To proceed further, we consider the citation data of top 10, 20, and 50 institutes or universities from each country across the globe. This data is found to be spread over 72, 55 and 25 countries, respectively  for top 10, 20 and 50 institutions. 
It is fascinating to note from Fig.~\ref{fig:pow} that the  power law behaviour holds for all these 3 separate cases as well.  
Each plot in Fig.~\ref{fig:pow} is accompanied with its respective KS distance $(KS)$ and  $p$-value for the fit as well as the corresponding power law exponent  $(\nu)$. 
However, there is a variation in the value of the exponent with the change in the number of institutions considered for the analysis. 
The adherence of the consistent pattern of power law in the higher end of the citation counts~\citep{Red_1998} indicates a predictable relationship between the rank of an institution and its citation count across different scales. 
Some recent studies have also demonstrated this power law trend in citation analyses \citep{pow1,pow2,pow3}.

\newpage
\subsubsection{Citation diversity in top institutes across the globe}
\begin{figure}[H]
    \centering
     \begin{subfigure}[b]{0.49\textwidth}
         \centering
         \includegraphics[width=\textwidth]{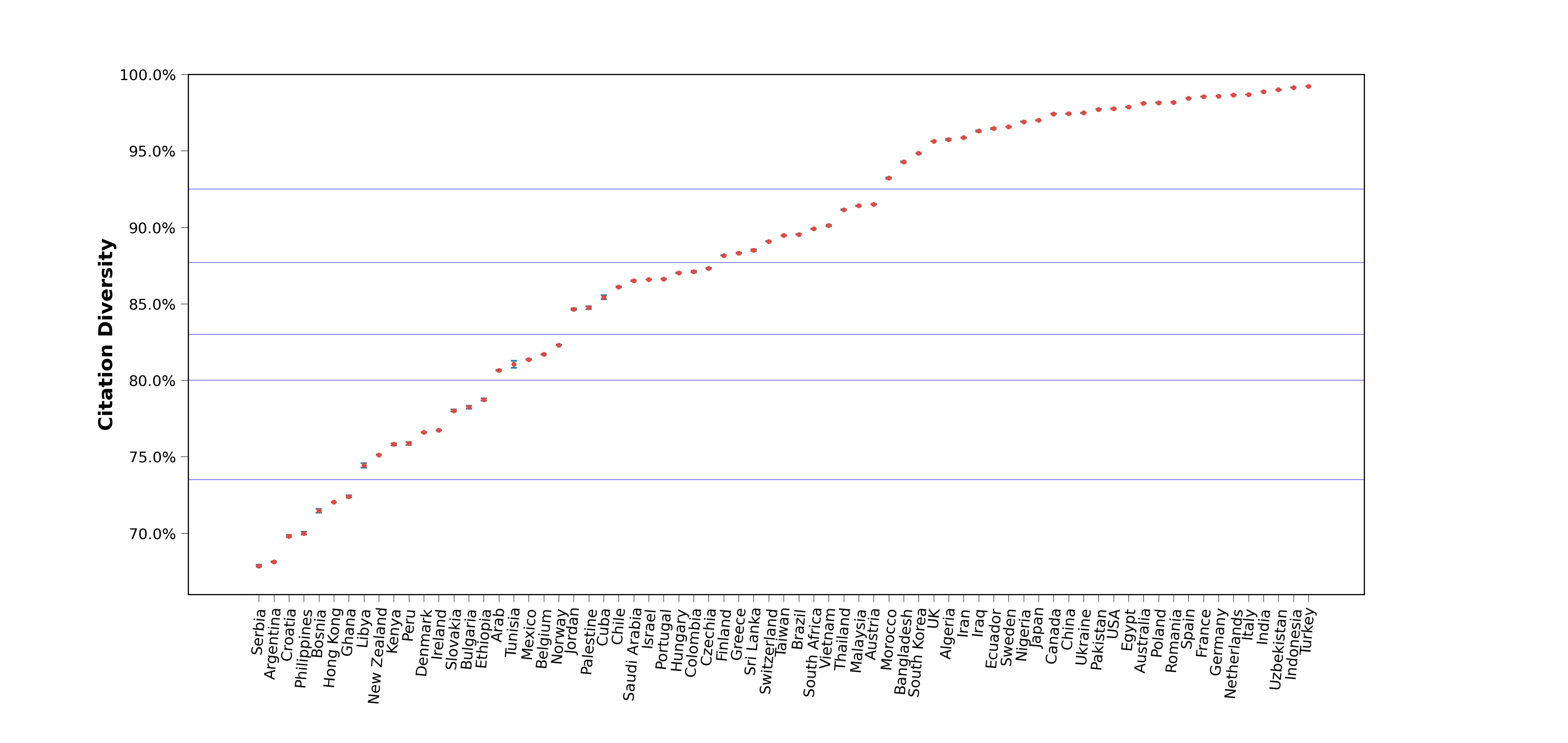}
         \caption{Citation diversity (D) with confidence intervals.}
         \label{fig:a}
     \end{subfigure}
     \hfill
     \begin{subfigure}[b]{0.49\textwidth}
         \centering
         \includegraphics[width=\textwidth]{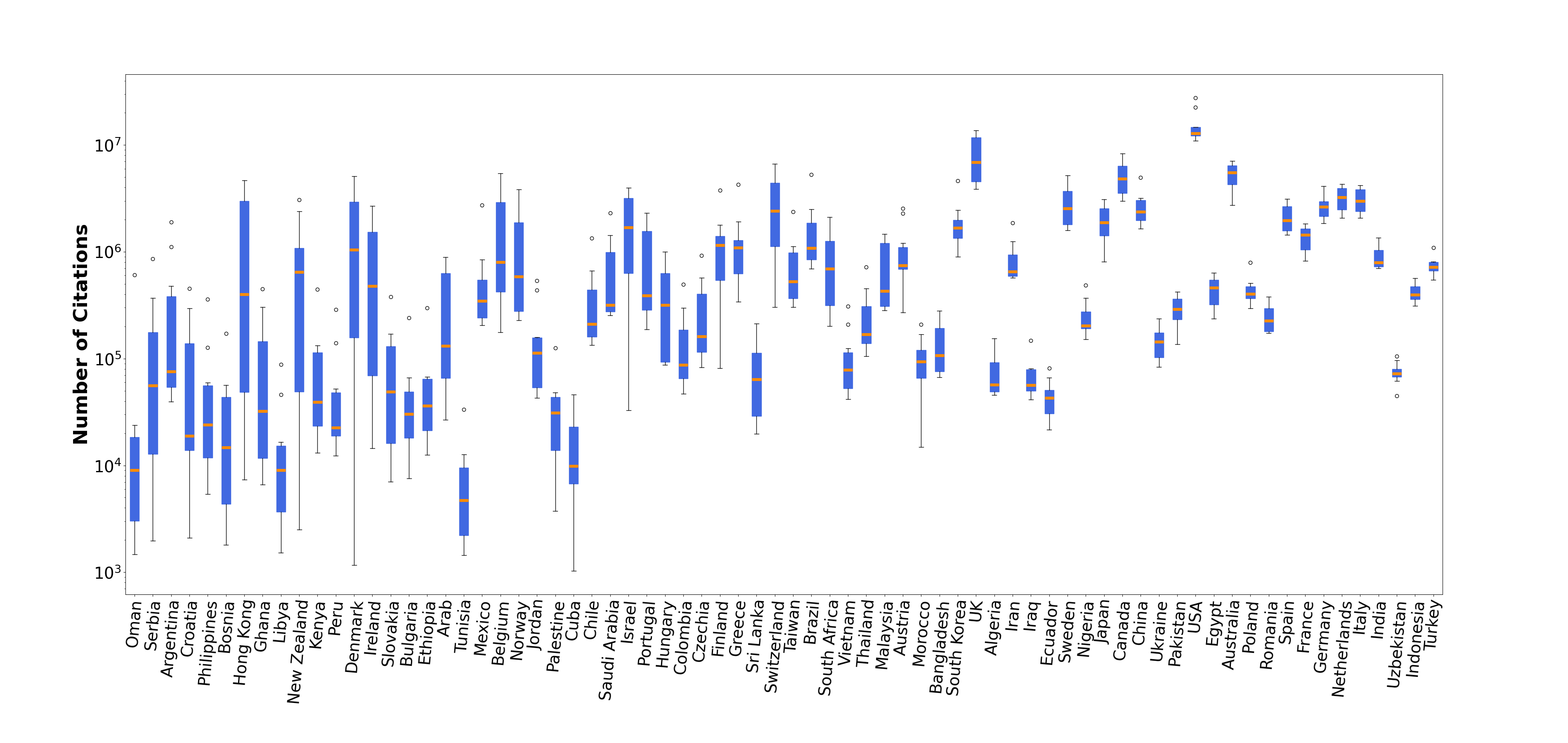}
         \caption{Box-plot of the total citation count (in log-scale).}
         \label{fig:b}
     \end{subfigure}
        \caption{(a) Citation diversity measure (red dot) for the top 10 institutes, along with their $95\%$ confidence intervals (blue vertical lines), for each country \footnotemark[1], 
        and (b) box-plot for the total citation of top 10 institutes across the globe. The countries are arranged in the same order of increasing values of D in both (a) and (b).}
        \label{fig:top 10}
\end{figure}

\begin{figure}[H]
     \centering
     \begin{subfigure}[b]{0.55\textwidth}
         \centering
         \includegraphics[width=\textwidth]{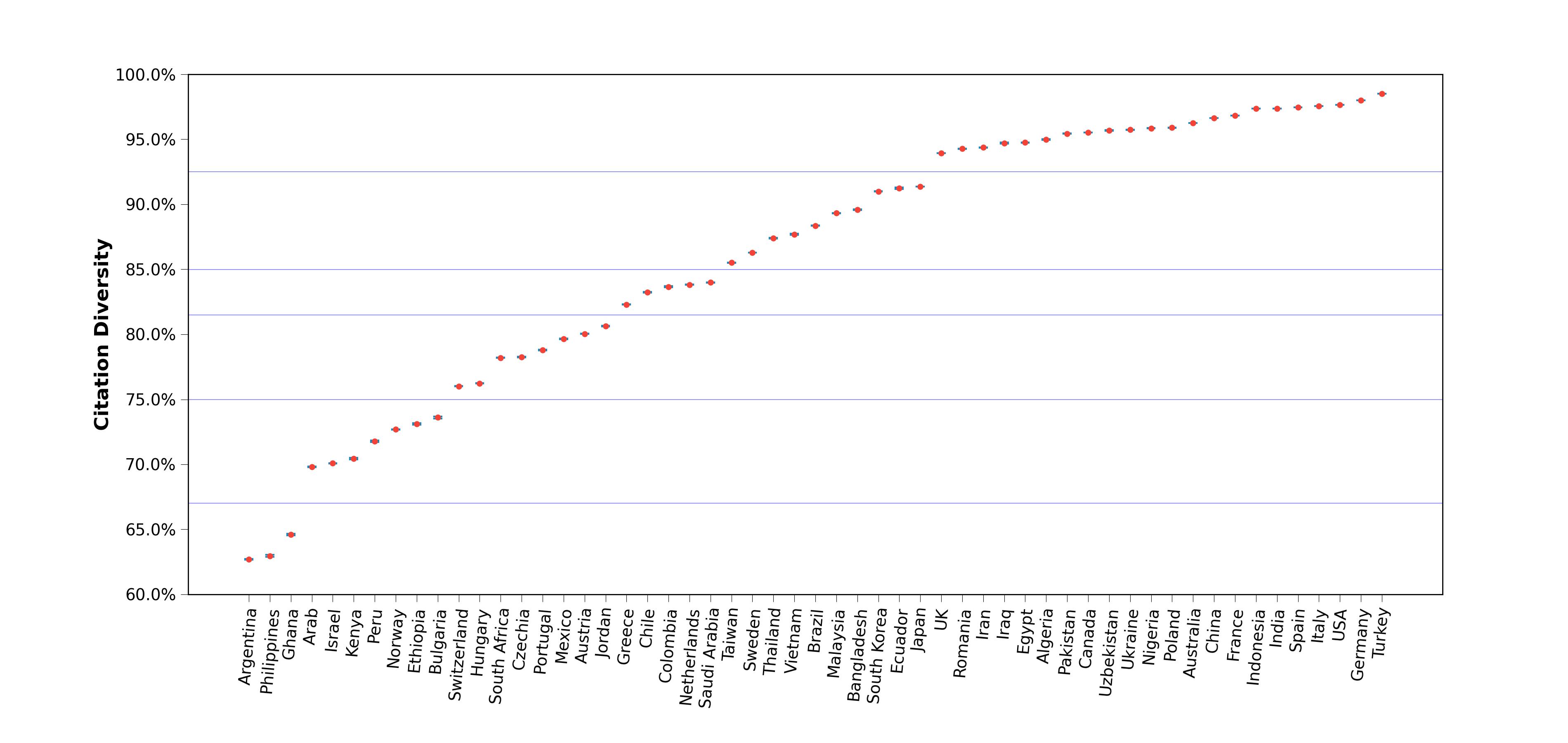}
         \caption{Citation diversity (D) with confidence intervals.}
         \label{fig:a1}
     \end{subfigure}
     \hfill
     \begin{subfigure}[b]{0.44\textwidth}
         \centering
         \includegraphics[width=\textwidth]{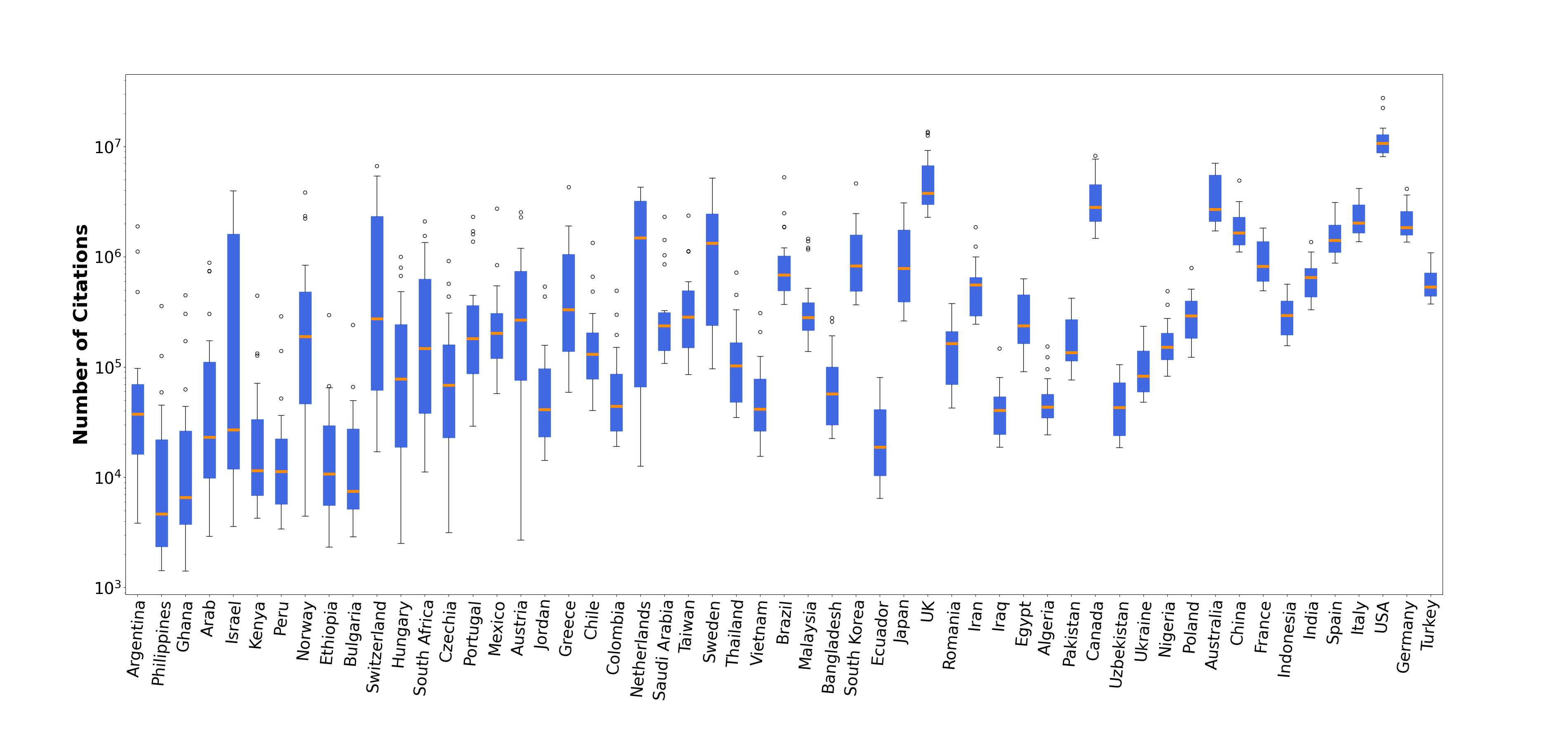}
         \caption{Box-plot of the total citation count (in log-scale).}
         \label{fig:b1}
     \end{subfigure}
        \caption{(a) Citation diversity measure (red dot) for the top 20 institutes, along with their $95\%$ confidence intervals (blue vertical lines), for each country and (b) box-plot for the total citation of top 20 institutes across the globe. The countries are arranged in the same order of increasing values of D in both (a) and (b).}
        \label{fig:top 20}
\end{figure}

\begin{figure}[H]
     \centering
     \begin{subfigure}[b]{0.55\textwidth}
         \centering
         \includegraphics[width=\textwidth]{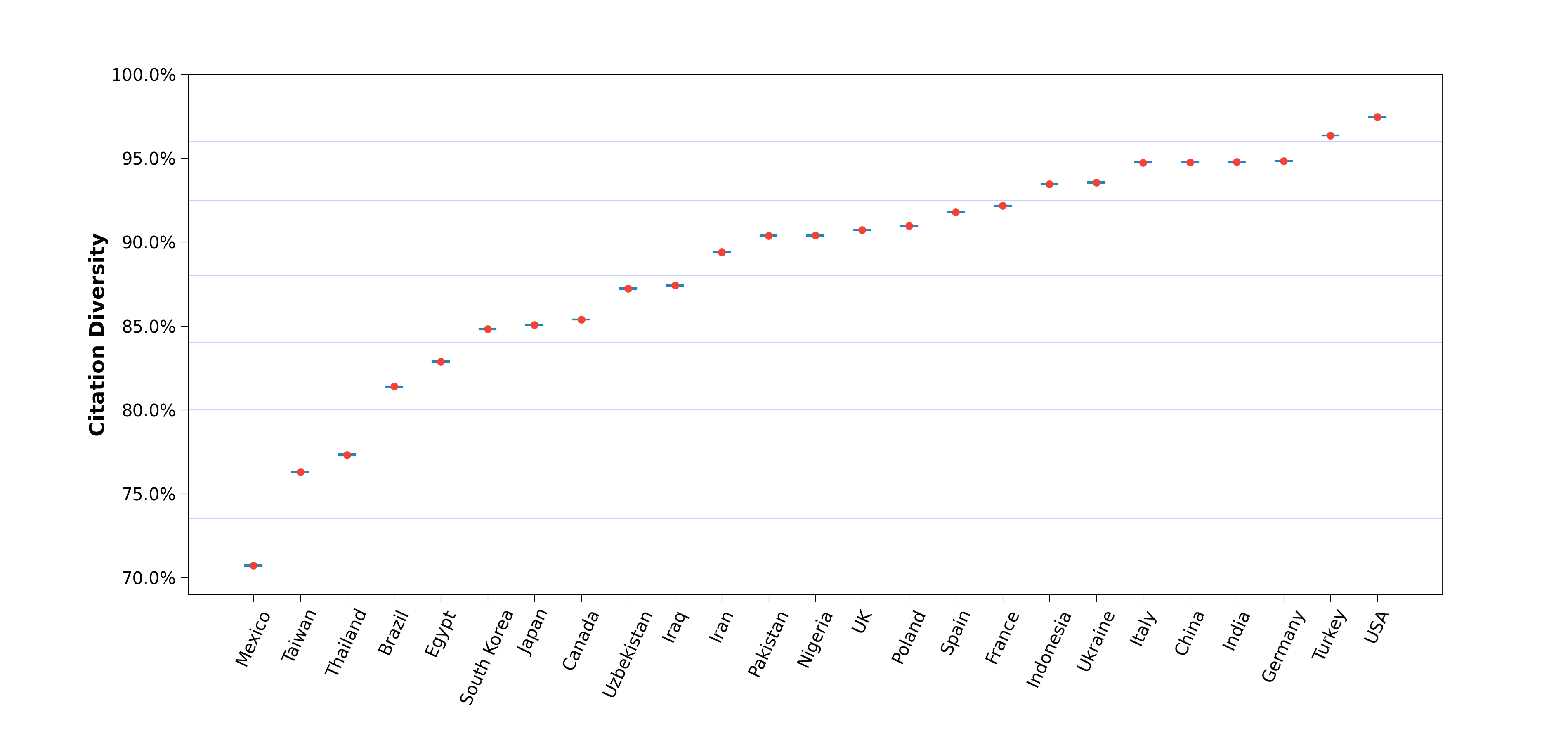}
         \caption{Citation diversity (D) with confidence intervals.}
         \label{fig:a2}
     \end{subfigure}
     \hfill
     \begin{subfigure}[b]{0.44\textwidth}
         \centering
         \includegraphics[width=\textwidth]{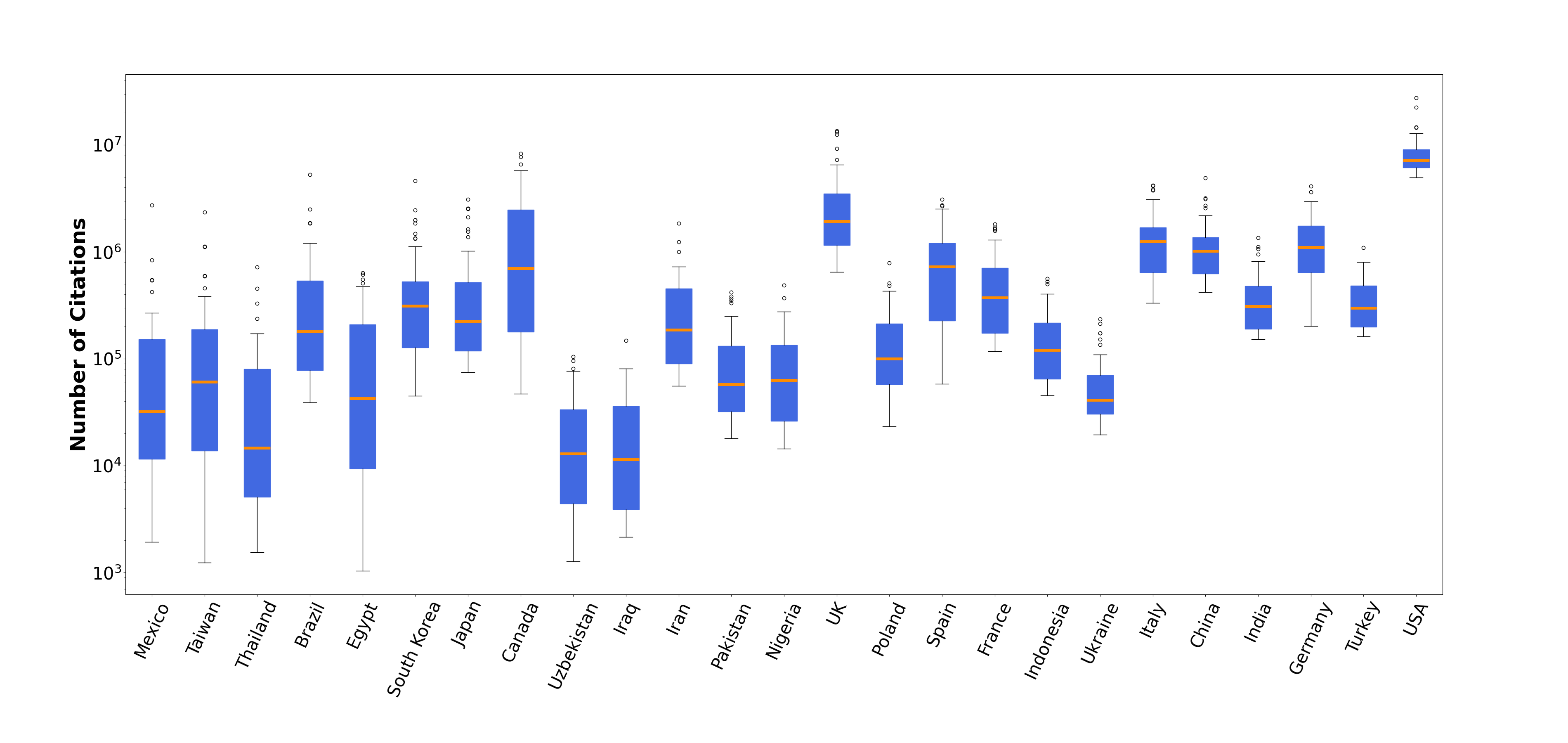}
         \caption{Box-plot of the total citation count (in log-scale).}
         \label{fig:b2}
     \end{subfigure}    
        \caption{(a) Citation diversity measure (red dot) for the top 50 institutes, along with their $95\%$ confidence intervals (blue vertical lines), for each country and (b) box-plot for the total citation of top 50 institutes across the globe. The countries are arranged in the same order of increasing values of D in both (a) and (b).}
        \label{fig:top 50}
\end{figure}

Next we have studied the diversity in the distribution of the citations of the top institutes or universities within each country. We employ the calculated D values, as previously explained in Section III, derived from the total citation count across all disciplines for each country's top 10, 20, and 50 institutions. This approach allows us to classify the countries based on these 
diversity values (D) and also into 3 subgroups within each group based on high, medium and low citation counts ($N_{c}$). Tables \ref{tab1},\ref{tab2} and \ref{tab3} provide detailed breakdown of these grouping, respectively, for the top 10, 20 and 50 institutions across various nations; associated confidence intervals of the diversity measures are presented in Figs.~\ref{fig:top 10},\ref{fig:top 20} and \ref{fig:top 50}, respectively, along with box-plot visualizations of raw total citation data in each cases. Given the wide range of $N_{c}$ counts per country, we use a logarithmic scale for the y-axis in our box-plots (Figs.~\ref{fig:b},\ref{fig:b1},\ref{fig:b2}) to effectively capture and represent the distribution of citation counts within different countries.

In general, we have noted that some countries, despite having a high $N_{c}$ count, do not necessarily have high D values. Conversely, there are countries with lower $N_{c}$ counts that exhibit very high D values. Therefore, a combined analysis offers insights into both the overall diversity and the spread of citations among top institutes/universities across various countries.
\paragraph{Results for top 10 institutions}
In this analysis of top 10 institutes across various countries, we examine 72 countries and divide them into six distinct groups (Group A - Group F) based on their decreasing diversity values (D), with each group being closely homogeneous in terms of their values of D.
Each of these groups are again divided into 3 subgroups based on high, medium, and low citation counts $N_{c}$ for each country (see Table \ref{tab1}). 
For example, Group A countries with very high D, can be sub-grouped into  A1, A2 and A3 group of countries with  high, medium and low $N_{c}$ counts respectively. 
Notably, Fig.~\ref{fig:a} highlights the remarkably small confidence interval for each country's citation diversity, signifying a high degree of certainty in our diversity estimates. 

It is evident from the $N_{c}$ count data of Table~\ref{tab1} that, within Group A, the USA stands out as the most highly cited country when examining its top 10 institutions. However, our analysis reveals a different leader in Group A, with Turkey emerging as the country with the highest citation diversity. 
In Group B, while Switzerland emerges as the most frequently cited country, our study shows that Austria exhibits the highest citation diversity within the group. This indicates that although Switzerland may dominate in terms of citation volume, Austria's citations are more evenly distributed among its top 10 institutes. 
Conversely, Finland, despite being a part of the highly cited subgroup B1, registers the lowest citation diversity in Group B, suggesting a more concentrated citation pattern. Meanwhile, in Group D, Belgium stands out as the most frequently cited country, yet Norway surpasses it in citation diversity, indicating a wider spread of citations across top Norwegian institutes/universities. 
In all other groups and subgroups similar kind of results can be inferred. It is apparent that relying solely on total citation value or average citation counts fails to adequately appreciate the impression of citation analysis; 
the citation diversity measures are also required for a complete picture. 
\begin{table}[H]
 \centering
 \small
\begin{tabular}{|c|c|l|r|c|}
\hline
~~~Group~~~ & Sub-Gr. & Country & $N_{c}$ & D \\
\hline
&\multirow{13}{*}{A1} &USA&15258270.00&97.86\\
\cline{3-5}
&&UK&8033507.00&95.74\\
\cline{3-5}
&&Australia&5196256.00&98.23\\
\cline{3-5}
&&Canada&5156112.00&97.53\\
\cline{3-5}
&&Netherlands&3205991.00&98.76\\
\cline{3-5}
&&Italy&3118657.00&98.79\\
\cline{3-5}
&&Sweden&2823378.00&96.69\\
\cline{3-5}
&&Germany&2718800.00&98.69\\
\cline{3-5}
&&China&2607126.00&97.54\\
\cline{3-5}
&&Spain&2117867.00&98.55\\
\cline{3-5}
&&Japan&1924270.00&97.10\\
\cline{3-5}
&&South Korea&1913191.00&94.94\\
\cline{3-5}
Gr. A &&France&1358776.00&98.65\\
\cline{2-5}
countries  & \multirow{11}{*}{A2}&India&897276.80& 98.98\\
\cline{3-5}
$D\in(93,99)$&&Iran&848094.30&95.97\\
\cline{3-5}
&&Turkey&742169.30&99.33\\
\cline{3-5}
&&Egypt&441066.70&97.98\\
\cline{3-5}
&&Poland&438283.70&98.25\\
\cline{3-5}
&&Indonesia&415959.30&99.25\\
\cline{3-5}
&&Pakistan&287339.10&97.83\\
\cline{3-5}
&&Nigeria&251443.80&97.02\\
\cline{3-5}
&&Romania&244935.50&98.29\\
\cline{3-5}
&&Ukraine&146194.00&97.59\\
\cline{3-5}
&&Bangladesh&142648.90&94.38\\
\cline{2-5}
&\multirow{5}{*}{A3} & Morocco & 98366.10 & 93.33\\
\cline{3-5}
&& Algeria & 75732.40 & 95.84\\
\cline{3-5}
&& Uzbekistan & 74695.20	& 99.10 \\
\cline{3-5}
&& Iraq & 68101.10 & 96.40  \\
\cline{3-5}
&& Ecuador & 43819.60& 96.56  \\
\hline 
\hline
&\multirow{5}{*}{B1} & Switzerland & 2883689.00 & 89.18 \\
\cline{3-5}
&&Brazil&1679322.00&89.64\\
\cline{3-5}
&&Greece&1306760.00&88.41\\
\cline{3-5}
&&Finland&1247269.00&88.27\\
\cline{3-5}
Gr. B&&Austria&1054539.00&91.61\\
\cline{2-5}
countries&\multirow{5}{*}{B2}& South Africa & 845386.70 & 90.00\\
\cline{3-5}
$D\in(88,92)$&&Taiwan&764616.10&89.57\\
\cline{3-5}
&& Malaysia & 729253.30 & 91.51\\
\cline{3-5}
&&Thailand&259906.40&91.24\\
\cline{3-5}
&& Vietnam & 108368.90 & 90.21 \\
\cline{2-5}
&\multirow{1}{*}{B3}&Sri Lanka&87499.50&88.61\\
\hline
\end{tabular}
\begin{tabular}{|c|c|l|r|c|}
  \hline
~~~Group~~~ & Sub-Gr. & Country & $N_{c}$ & D\\
\hline
&\multirow{1}{*}{C1}&Israel&1897026.00&86.68\\
\cline{2-5}
&\multirow{7}{*}{C2}& Portugal & 880957.10 & 86.72\\
\cline{3-5}
&&Saudi Arabia&732839.40&86.61\\
\cline{3-5}
Gr. C&&Hungary&396155.60&87.13\\
\cline{3-5}
countries&&Chile&383313.90&86.19\\
\cline{3-5}
$D\in(85,88)$&&Czechia&299917.30&87.42\\
\cline{3-5}
&&Jordan&171519.80&84.74\\
\cline{3-5}
&&Colombia&154868.40&87.20\\
\cline{2-5}
&\multirow{2}{*}{C3}& Palestine & 35691.80 & 84.84\\
\cline{3-5}
&&Cuba&16006.10&85.54\\
\hline
\hline
&\multirow{2}{*}{D1}& Belgium & 1819939.00 & 81.79\\
\cline{3-5}
Gr. D&&Norway&1149549.00&82.40\\
\cline{2-5}
countries&\multirow{2}{*}{D2}&Mexico&627030.00&81.44\\
\cline{3-5}
$D\in(80,83)$&&Arab&314101.00&80.75\\
\cline{2-5}
&\multirow{1}{*}{D3}& Tunisia & 8102.40 & 81.14\\
\hline
\hline
&\multirow{1}{*}{E1} &Denmark&1631915.00&76.68\\
\cline{2-5}
&\multirow{2}{*}{E2} & New Zealand & 906472.50 & 75.22 \\
\cline{3-5}
& &Ireland & 893135.80 & 76.82 \\
\cline{2-5}
Gr. E&\multirow{6}{*}{E3} & Kenya & 93166.60 & 75.90\\
\cline{3-5}
countries&&Slovakia&92479.90&78.11\\
\cline{3-5}
$D\in(75,79)$&&Ethiopia&63702.60&78.83\\
\cline{3-5}
&&Peru&62510.60&75.95\\
\cline{3-5}
&&Bulgaria&51697.10&78.33\\
\cline{3-5}
&&Libya&19220.90&74.52\\
\hline
\hline
&\multirow{1}{*}{F1}&Hong Kong&1570705.00&72.12\\
\cline{2-5}
&\multirow{4}{*}{F2}&Argentina&394459.70&68.21\\
\cline{3-5}
Gr. F&&Serbia&170841.90&67.95\\
\cline{3-5}
countries&&Ghana&109620.90&72.45\\
\cline{3-5}
$D\in(67,73)$&&Croatia&105208.20&69.88\\
\cline{2-5}
&\multirow{2}{*}{F3}&Philippines&67549.00&70.09\\
\cline{3-5}
&&Bosnia&34852.90&71.55\\
\hline
\hline
Outlier&&Oman&69586.90&41.06\\
\hline
\multicolumn{5}{c}{}\\
\multicolumn{5}{c}{}\\
\multicolumn{5}{c}{}\\
\multicolumn{5}{c}{}\\
\multicolumn{5}{c}{}\\
\multicolumn{5}{c}{}\\
\multicolumn{5}{c}{}\\
\end{tabular}
\caption{Categorization of countries (based on the top 10 institutes or universities) into groups and sub-groups according to diversity values (D) and average citation count ($N_{c}$) per institute.}
\label{tab1}
\end{table}

\newpage
\footnotetext[1]{excluding Oman due to its significantly lower value compared to other countries}
\paragraph{Results for top 20 institutes} 
By broadening our analysis to include the top 20 institutes, we have been able to study 55 countries across the globe as per the availability of data. In this case, we observe significant changes, compared to top 10 institutions, both in the diversity measure (D) and the average citation count ($N_{c}$)  for each country. Countries are again grouped as per their values of D and $N_{c}$ as in the case of top 10 institutes (Table \ref{tab2} and Fig.~\ref{fig:top 20}). This expanded view makes the distinctions between countries more apparent. For instance, Israel is initially ranked in Group C with high D and maximum $N_{c}$ within this group when considering the top 10 institutes. However, when the scope is broadened to include the top 20 institutes, its performance metrics decline, moving it to Group E with a significantly lower D value. Similarly, Netherlands is categorized in Group C with lower D and $N_{c}$ values when examining the top 20 institutes, but rises to Group A with much higher D and $N_{c}$ values when focusing on the top 10 institutes. 
These results imply that, in Finland and Netherlands, institutes ranked within 11 to 20 have significantly diverse and have lower citation counts compared to the top 10 institutes there, which were much more homogeneous in terms of citation counts.
In contrast, while considering the top 10 institutes, Morocco is positioned in Group A with a much higher D and $N_{c}$ values, but completely drops out of the rankings when the scope is expanded to the top 20 institutes (as their is not many institutes outside the top 10 list in Morocco to have sizable/reportable citation data).

\begin{table}[H]
 \centering
 \small
\begin{tabular}{|c|c|l|r|c|}
\hline
~~~Group~~~ & Sub-Gr. & Country & $N_{c}$ & D \\
\hline
&\multirow{8}{*}{A1} &USA&12105340.00&97.63\\
\cline{3-5}
 &&UK	&5523985.00&	93.92\\
 \cline{3-5}
 &&Australia&	3715297.00&	96.24\\
 \cline{3-5}
&& Canada&	3638187.00	&95.51\\
\cline{3-5}
&& Italy&	2367446.00&	97.54\\
\cline{3-5}
&& Germany	&2143908.00&97.98\\
\cline{3-5}
&& China&1955727.00&96.63\\
\cline{3-5}
&&Spain	&1613509.00&97.44\\
 \cline{2-5}
& \multirow{11}{*}{A2}&France&993184.30&96.82\\
\cline{3-5}
 Gr. A&&India&671044.80&97.36\\
 \cline{3-5}
countries &&Iran&599467.90&94.36\\
 \cline{3-5}
$D\in(93,99)$ &&Turkey&594240.60&98.50\\
 \cline{3-5}
 &&Poland&310970.30&95.89\\
 \cline{3-5}
 &&Indonesia	&307798.30&	97.35\\
 \cline{3-5}
 &&Egypt&300285.20&94.75\\
 \cline{3-5}
 &&Pakistan	&199903.00&	95.42\\
 \cline{3-5}
 &&Nigeria&	183957.10&	95.84\\
 \cline{3-5}
 &&Romania&	164146.60&	94.27\\
 \cline{3-5}
 &&Ukraine&	104113.40&	95.73\\
 \cline{2-5}
 &\multirow{3}{*}{A3}&Algeria&54751.30&94.98\\
 \cline{3-5}
 &&Uzbekistan&	51147.85&	95.67\\
 \cline{3-5}
 &&Iraq&	48109.00&	94.70\\
\hline
\hline
&\multirow{4}{*}{B1} &Sweden&	1570605.00&	86.28\\
\cline{3-5}
 &&South Korea	&1205591.00&	90.98\\
 \cline{3-5}
 &&Japan&	1181852.00&	91.35\\
 \cline{3-5}
&& Brazil&	1086417.00&	88.35\\
\cline{2-5}
Gr. B&\multirow{3}{*}{B2}&Malaysia	&466856.50&	89.32\\
\cline{3-5}
countries &&Taiwan	&458992.30&	85.50\\
 \cline{3-5}
$D\in(85,92)$ &&Thailand	&159469.00	&87.40\\
 \cline{2-5}
&\multirow{3}{*}{B3}&Bangladesh	&87134.20&	89.57\\
 \cline{3-5}
 &&Vietnam&	68064.05	&87.69\\
 \cline{3-5}
 &&Ecuador&	26947.45&	91.23\\
 \hline
\end{tabular}
\begin{tabular}{|c|c|l|r|c|}
  \hline
~~~Group~~~ & Sub-Gr. & Country & $N_{c}$ & D \\
\hline 
&\multirow{1}{*}{C1}&Netherlands&	1699040.00	&83.82\\
\cline{2-5}
Gr. C&\multirow{3}{*}{C2}& Greece & 743842.40 & 82.29\\
\cline{3-5}
countries&&Saudi Arabia&441429.00&83.99\\
\cline{3-5}
$D\in(82,84)$ &&Chile&	230567.20&	83.22\\
\cline{2-5}
&\multirow{1}{*}{C3}& Colombia&	91812.10&	83.66\\
\hline
\hline
&\multirow{1}{*}{D1}&Switzerland&1491331.00&76.01\\
\cline{2-5}
&\multirow{6}{*}{D2}&Austria&569110.70&80.03\\
\cline{3-5}
Gr. D&& Portugal	&480006.50&	78.79\\
\cline{3-5}
countries&& South Africa	&444548.00&	78.19\\
\cline{3-5}
$D\in(76,81)$ &&  Mexico	&372942.90&	79.65\\
\cline{3-5}
&& Hungary	&208885.20&	76.23\\
\cline{3-5}
&& Czechia&	162180.00&	78.25\\
\cline{2-5}
&\multirow{1}{*}{D3}& Jordan&98843.60&	80.62\\
\hline
\hline
&\multirow{3}{*}{E1} &Israel&954046.30&	70.07\\
\cline{3-5}
 &&Norway&	606315.00&	72.68\\
 \cline{3-5}
Gr. E &&Arab&	161761.90&	69.81\\
\cline{2-5}
countries&\multirow{4}{*}{E2} &  Kenya&50018.80&70.45\\
\cline{3-5}
$D\in(69,74)$ &&Ethiopia&	34585.35&73.11\\
 \cline{3-5}
 &&Peru	&34206.85&71.77\\
 \cline{3-5}
 &&Bulgaria&	28378.05&73.60\\
\hline
\hline
Gr. F&\multirow{1}{*}{F1}&Argentina&206299.20&62.71\\
\cline{2-5}
countries&\multirow{2}{*}{F2}&Ghana&	56659.70&64.61\\
\cline{3-5}
$D\in(62,65)$ &&Philippines&	34953.20&	62.96\\
\hline
\multicolumn{5}{c}{}\\
\multicolumn{5}{c}{}\\
\multicolumn{5}{c}{}\\
\multicolumn{5}{c}{}\\
\multicolumn{5}{c}{}\\
\multicolumn{5}{c}{}\\
\multicolumn{5}{c}{}\\
\multicolumn{5}{c}{}\\
\end{tabular}
\caption{Categorization of countries (based on the top 20 institutes or universities) into groups and sub-groups according to diversity values (D) and average citation count ($N_{c}$) per institute.}
\label{tab2}
\end{table}
\paragraph{Results for top 50 institutes} 
When we focus on the citation data for top 50 institutes, we get data only on
25 countries whose diversity values are calculated from their total citations (Table \ref{tab3} and Fig.~\ref{fig:top 50}). 
In the analysis, focusing on the top 50 institutions, Taiwan and Thailand fall into Group E, characterized by lower D values. However, when considering only the top 20 institutions in these countries, they move to Group B, which has comparatively higher D values. Conversely, Spain is placed in Group A with high D and $N_{c}$ values when considering the top 20 institutions, but it shifts to Group B with lower D and $N_{c}$ values when the top 50 institutions are considered. In Group C, Canada is grouped with South Korea, Japan, and others, sharing similar D values but with significantly different $N_{c}$ values when considering the top 50 institutions. However, when focusing on the top 20 institutions, Canada moves to Group A, while South Korea and Japan are in Group B, with different D values.

\begin{table}[H]
 \centering
  \small
\resizebox{\columnwidth}{!}{%
\begin{tabular}{|c|c|l|r|c|}
\hline
~~~~~~Group~~~~~~ & Sub-Gr. & Country & $N_{c}$ & D \\
\hline
&\multirow{4}{*}{A1} &USA&8666838.00&97.52\\
\cline{3-5}
&& Italy&1431165.00&	94.79\\
 \cline{3-5}
Gr. A &&Germany&	1303227.00	&94.89\\
 \cline{3-5}
countries && China&	1211080.00&	94.82\\
\cline{2-5}
$D\in (93,98)$ &\multirow{3}{*}{A2}&India	&402849.20	&94.82\\
\cline{3-5}
&&Turkey&378319.00 &96.41\\
\cline{3-5}
 &&Indonesia	&173539.30&	93.51\\
 \cline{2-5}
&\multirow{1}{*}{A3}&Ukraine&	61362.62	&93.61\\
 \hline
 \hline
&\multirow{1}{*}{B1}& UK&	3022509.00&	90.76\\
\cline{2-5}
&\multirow{5}{*}{B2}& Spain	&883924.30	&91.84\\
 \cline{3-5}
 Gr. B&&France&	539431.30&	92.22\\
 \cline{3-5}
countries &&Iran&	309750.10&	89.44\\
 \cline{3-5}
$D\in(89,93)$ &&Poland&	164158.20&	91.01\\
 \cline{3-5}
 &&Pakistan	&104250.40	&90.43\\
 \cline{2-5}
&\multirow{1}{*}{B3}& Nigeria&	96024.60&	90.45\\
\hline
\end{tabular}
\begin{tabular}{|c|c|l|r|c|}
\hline
~~~~~~Group~~~~~~ & Sub-Gr. & Country & $N_{c}$ & D \\
\hline
&\multirow{1}{*}{C1}& Canada	&1675403.00&85.43\\
\cline{2-5}
Gr. C&\multirow{2}{*}{C2}& South Korea&	581050.20&84.85\\
\cline{3-5}
countries &&Japan&	562057.30&	85.12\\
\cline{2-5}
$D\in(84,88)$&\multirow{2}{*}{C3}& Uzbekistan&	24379.08&87.26\\
\cline{3-5}
&&Iraq&23362.16&87.47\\
\hline
\hline
Gr. D countries&\multirow{1}{*}{D1}& Brazil&	501581.10&	81.43\\
\cline{2-5}
$D\in(81,83)$&\multirow{1}{*}{D2}&Egypt	&132562.70	&82.92\\
\hline 
\hline 
Gr. E &\multirow{2}{*}{E1}&  Taiwan&	200544.00&	76.34\\
\cline{3-5}
countries &&Mexico	&158899.30&	70.76\\
\cline{2-5}
$D\in(70,78)$&\multirow{1}{*}{E2}& Thailand &68940.40&	77.37\\
\hline
\multicolumn{5}{c}{}\\
\multicolumn{5}{c}{}\\
\multicolumn{5}{c}{}\\
\multicolumn{5}{c}{}\\
\end{tabular}
}
\caption{Categorization of countries (based on the top 50 institutes or universities) into groups and sub-groups according to citation diversity values (D) and average citation count $(N_{c})$ per institute.}
\label{tab3}
\end{table}


In conclusion, our citation diversity metric complements total citation counts by providing additional insights that cannot be captured by citation counts alone. While total citations reflect overall research output and impact, the diversity metric highlights the evenness of the distribution of citations across different institutions, disciplines, and demographics. By considering both measures together, we gain a more comprehensive and nuanced understanding of a nation's research landscape. In particular, this diversity index D ranges from 0 to 100 and can be interpreted as follows: a score of 100 would signify a perfectly even distribution of citations, while a score near 0 would indicate a highly uneven distribution. Therefore, for an example, Switzerland’s score of $D = 89.18$ suggests that while the distribution is quite balanced, there is still some degree of unevenness which may be improved further (this unevenness is indeed more compared to the countries in Group A having values of $D > 93$). The value of D are thus used for a basis of comparisons between countries in Tables~\ref{tab1},\ref{tab2} and \ref{tab3}, where countries in Group A have the most evenly distributed citations, while Group B includes countries with slightly less even citation distributions among their top institutes. The subsequent groups display progressively lower levels of evenness as the diversity scores decrease.
We observed significant variations in diversity depending on different numbers of top institutes. For instance, Israel's diversity decreased from $86.68\%$ (Group C) while considering top 10 institute to $70.07\%$ (Group E) while considering top 20 institute; the country is even dropped out of the top 50 list entirely, suggesting that the country has fewer than 50 renowned institutes, and the top 10 institutes are more homogeneous in terms of citation counts than the top 20 institutes there. This highlights the crucial influence of a country's concentration of high-performing institutes on its overall diversity score. Moreover, we observe that while the UK exhibits very high total citations across its top 10, 20, and 50 institutes, it is only classified in group A, characterized by a diversity range of 93\% to 99\%, when considering its top 10 and 20 institutes. However, when the top 50 institutes are taken into account, despite the high citation counts, the diversity value decreases, placing the UK in group B, with a diversity range of 89\% to 93\%. This indicates that although the UK maintains a strong citation performance, the citation diversity varies significantly with the number of institutes considered. When focusing on a smaller number of top institutes, the UK demonstrates a broader citation diversity, suggesting a wide-reaching influence of its most prominent research institutions. Conversely, India and USA displayed remarkable consistency in its diversity across all three institute tiers, suggesting a more balanced distribution of citations. However, expanding the scope to include more institutes reveals a drop in diversity, implying a more concentrated citation pattern. This highlights the importance of considering both citation count and diversity to fully understand the impact and reach of a country’s research output across different academic institutions across the globe. Total citation counts often mask the underlying distribution of citations, potentially misleading interpretations. By employing our novel metric, we gain a clearer picture of how citations are distributed across a country's top research institutions.  Our approach provides a more nuanced understanding of a nation's research landscape by revealing the distribution of citations amongst its leading institutions. 

\subsection{Citation diversity in various scientific disciplines}
Our citation diversity analysis in the previous section has been performed at the institutional level, irrespective of individual scientists or any specific scientific discipline. We now shift our focus to study the  citation diversity in the publication data of various scientific disciplines. We specifically explore the citation data of 126 internationally acclaimed elite researchers, in six important disciplines; physics, chemistry, mathematics, computer science, economics and physiology/medicine, 
21 from each discipline. 
Additionally, to see whether the citation pattern in various scientific disciplines has changed over recent times or not, we also explored the citation data of a total of 63 Nobel prize winners in physics, chemistry and physiology/medicine.



\subsubsection{Award winning scientists in recent times}
\begin{figure}[h]
     \centering
     \begin{subfigure}[b]{0.32\textwidth}
         \centering
         \includegraphics[width=\textwidth]{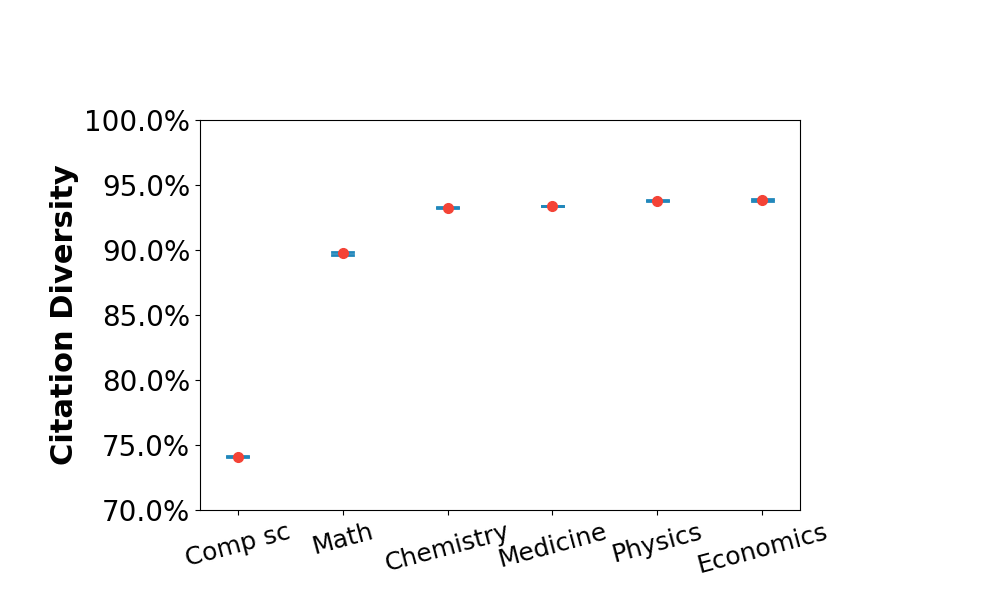}
         \caption{D for total citation.}
         \label{fig:5a}
     \end{subfigure}
     \hfill
     \begin{subfigure}[b]{0.32\textwidth}
         \centering
         \includegraphics[width=\textwidth]{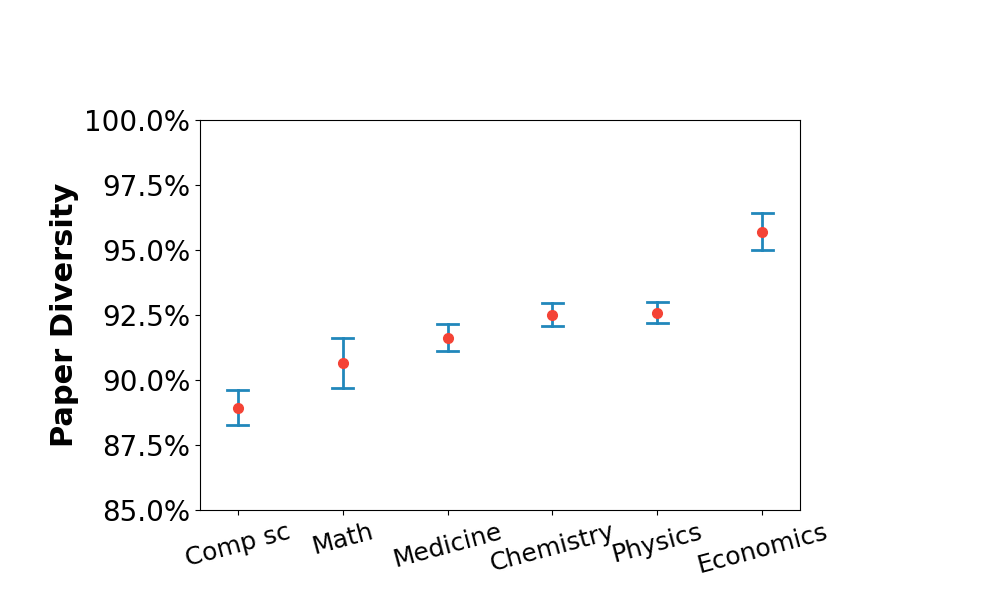}
         \caption{D for total paper.}
         \label{fig:5b}
     \end{subfigure}
     \hfill
     \begin{subfigure}[b]{0.32\textwidth}
         \centering
         \includegraphics[width=\textwidth]{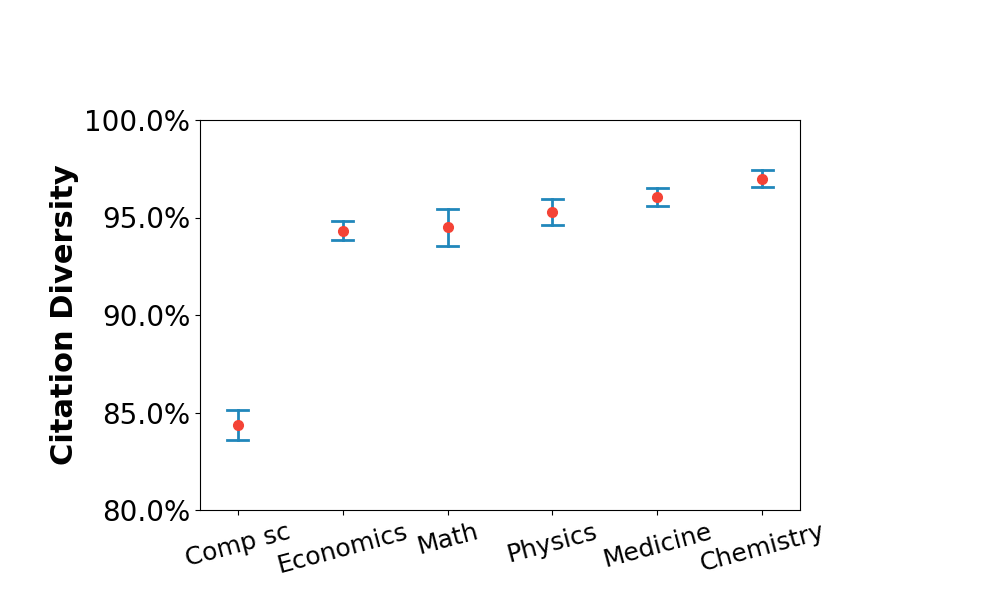}
         \caption{D for per-paper citation.}
         \label{fig:5c}
     \end{subfigure}
      \hfill
     \begin{subfigure}[b]{0.32\textwidth}
         \centering
         \includegraphics[width=\textwidth]{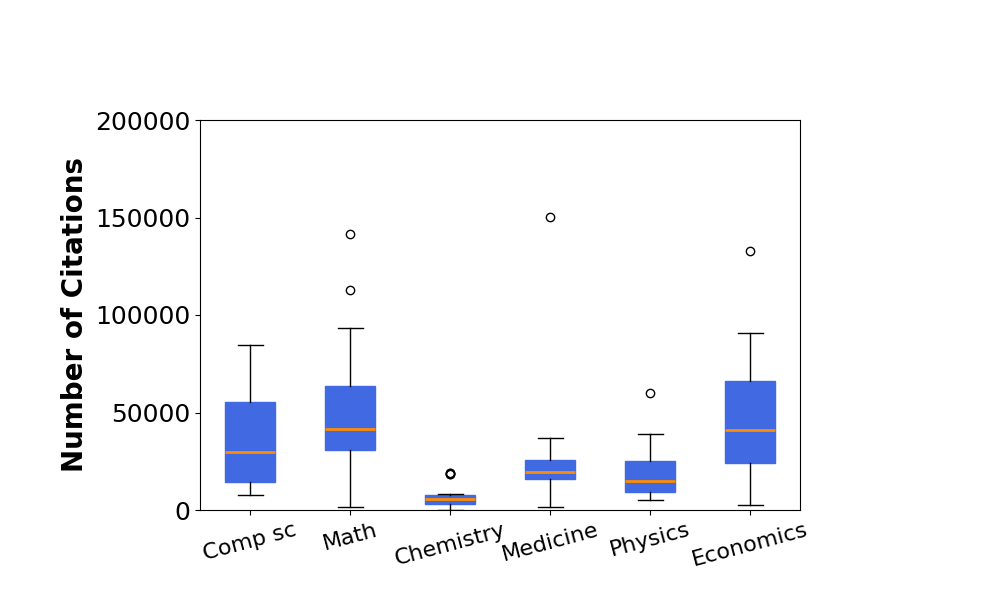}
         \caption{Box-plot of total citations.}
         \label{fig:5d}
     \end{subfigure}
      \hfill
     \begin{subfigure}[b]{0.32\textwidth}
         \centering
         \includegraphics[width=\textwidth]{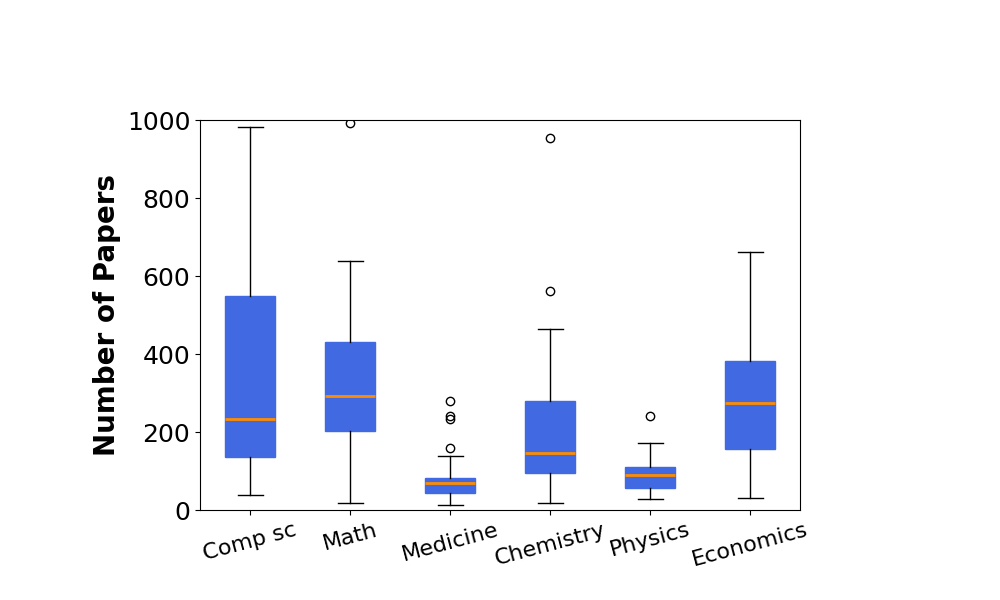}
         \caption{Box-plot of total papers.}
         \label{fig:5e}
     \end{subfigure}
      \hfill
     \begin{subfigure}[b]{0.32\textwidth}
         \centering
         \includegraphics[width=\textwidth]{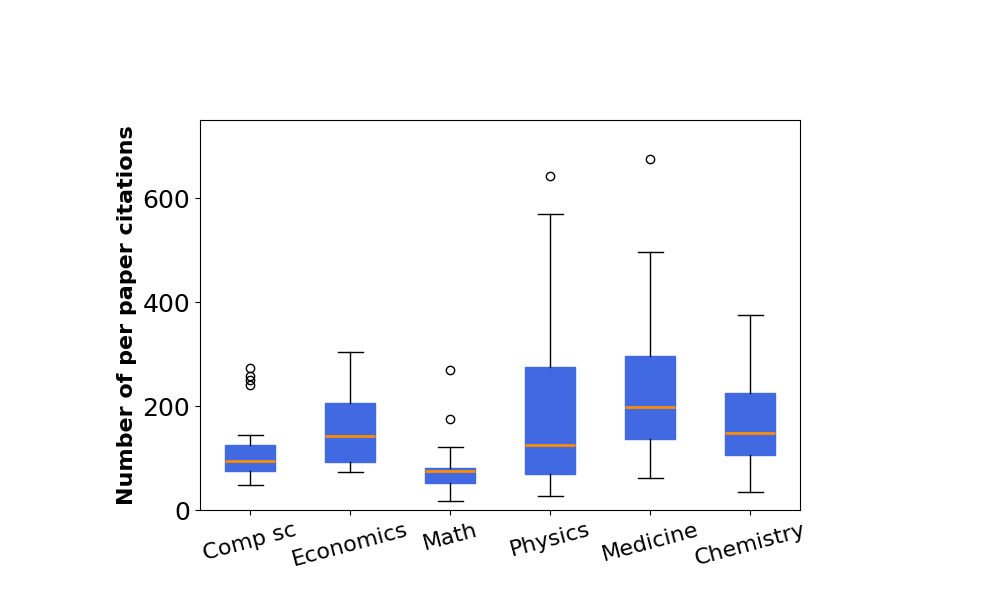}
         \caption{Box-plot of per-paper citation.}
         \label{fig:5f}
     \end{subfigure}
        \caption{(a),(b) and (c) represent the citation diversity measures (D) of recent award winners with red dots, along with their 95\% confidence intervals (blue vertical lines), for each cases. Also (d),(e) and (f) present the corresponding box-plots of the data.}  \label{fig:award_recent}
\end{figure}

To develop a thorough understanding of citation diversity in different scientific disciplines, we implement our methodology from three distinct viewpoints, namely total citations, total publications and per-paper citation count.
Table~\ref{tab4} showcases the calculated D values and $N_{c}$ counts per scientist for every discipline and  
Fig.~\ref{fig:award_recent} depicts these diversity measures, along with their 95\% confidence intervals, and the distributions of individual numbers through their box-plots.

\begin{table}[h]
 \centering
 \small
\begin{tabular}{|l|r|c|r|c|r|c|}
\cline{1-7}
 & \multicolumn{2}{c|} {Total Citation} & \multicolumn{2}{c|} {Total Publication}  & \multicolumn{2}{c|} {Per-paper Citation} \\
\cline{2-7}
\shortstack[lb]{Disciplines\\~} & $N_{c}$ & D & $N_{c}$ & D & $N_{c}$ & D \\
\cline{1-7}
Physics (2017-2023) &35982.52 &93.91 &364.71&92.73 & 121.00	&95.41\\
\cline{1-7}
Chemistry (2015-2023) &50636.52 &93.38 & 366.90&92.65& 153.70&	97.14\\
 \cline{1-7}
Mathematics (2007-2023) &7117.48 & 89.89& 89.62 &90.80 & 82.48&	94.63\\
\cline{1-7}
 Computer Science (2010-2023)&54503.90 &74.22 & 228.95&89.06 & 236.53&	84.47\\
\cline{1-7}
 Economics (2013-2023)&19426.90 &93.97 & 92.10	&95.85 & 240.22&	94.47\\
\cline{1-7}
Physiology/Medicine (2014-2023)& 48794.19	&93.52 & 330.57	&91.77 & 165.82&	96.20\\
\hline 
\end{tabular}
\caption{Average citation count ($N_{c}$) per scientist (considering 21 scientists in each discipline) and diversity value (D) for total citation, total publication, and per-paper citation across six scientific disciplines in recent times.}
\label{tab4}
\end{table}

\noindent It is noted that $N_{c}$ count per scientist in mathematics is minimum whereas its D value is not so low. On the other hand, the $N_{c}$ count per scientist in computer science is maximum but its D value is the lowest. So we can say that the total citation count of award winners in computer science is very high as compared to the other disciplines, but the diversity of papers in computer science is relatively low compared to other subjects. This indicated that, in computer science some award winners have excessively high number of papers and citations compared to some others. However, the number of papers and citation counts seems to be much more homogeneous across all winners in the other disciplines than in computer science. Additionally, the difference in paper and citation diversity among these other subjects is not as significant as the difference observed between computer science and them.

\subsubsection{Award winning scientists in old times in three principal disciplines}
\begin{figure}[h]
     \begin{subfigure}[b]{0.25\textwidth}
         \centering
         \includegraphics[width=\textwidth]{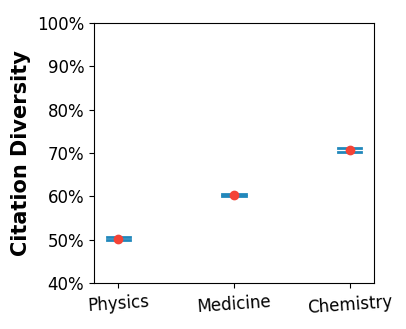}
         \caption{D for total citation.}
         \label{fig:6a}
     \end{subfigure}
     \hfill
     \begin{subfigure}[b]{0.25\textwidth}
         \centering
         \includegraphics[width=\textwidth]{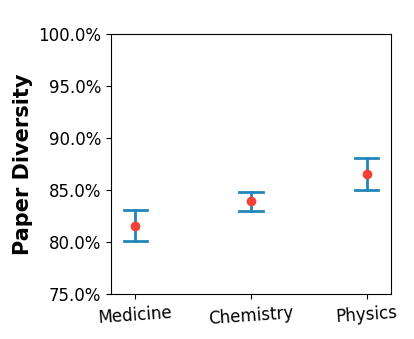}
         \caption{D for total publication.}
         \label{fig:6b}
     \end{subfigure}
     \hfill
     \begin{subfigure}[b]{0.25\textwidth}
         \centering
         \includegraphics[width=\textwidth]{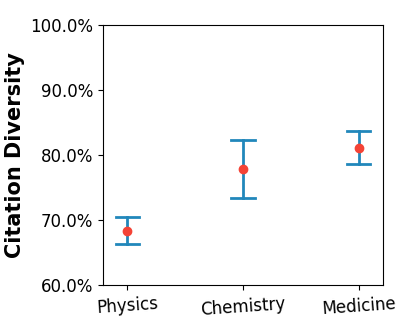}
         \caption{D for per-paper citation.}
         \label{fig:6c}
     \end{subfigure}
     \hfill
     \begin{subfigure}[b]{0.25\textwidth}
         \centering
         \includegraphics[width=\textwidth]{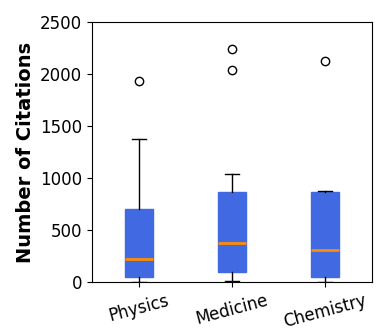}
         \caption{Box-plot of total citation.}
         \label{fig:6d}
     \end{subfigure}
     \hfill
     \begin{subfigure}[b]{0.25\textwidth}
         \centering
         \includegraphics[width=\textwidth]{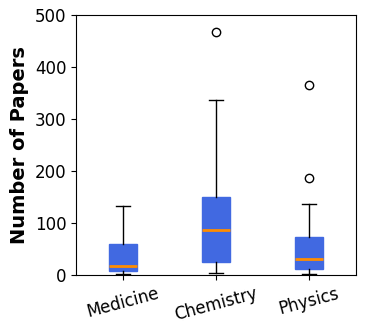}
         \caption{Box-plot of total papers.}
         \label{fig:6e}
     \end{subfigure}
     \hfill
     \begin{subfigure}[b]{0.25\textwidth}
         \centering
         \includegraphics[width=\textwidth]{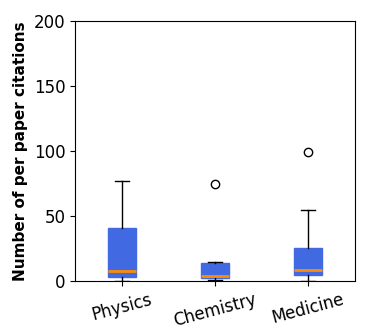}
         \caption{Box-plot of per-paper citation.}
         \label{fig:6f}
     \end{subfigure}
        \caption{(a),(b) and (c) represent the  diversity (D) of historical Nobel winners with red dots, along with their 95\% confidence intervals (blue vertical lines), for each cases. Also (d), (e) and (f) present the corresponding box-plots of the citation data.}
        \label{fig:award_old}
\end{figure}

We now extend our analysis to examine the citation diversity in the publication of century old Nobel winning scientists in physics, chemistry and physiology/medicine. 
We employ the three aforementioned viewpoints to calculate diversity percentage values for each discipline across historical periods (Table \ref{tab5}). 
Fig.~\ref{fig:award_old} illustrates their citation diversity and count distributions (box-plots).

\begin{table}[h]
 \centering
 \small
\begin{tabular}{|l|r|c|r|c|r|c|}
\cline{1-7}
 & \multicolumn{2}{c|} {Total Citation} & \multicolumn{2}{c|} {Total Publication}  & \multicolumn{2}{c|} {Per-paper Citation} \\
\cline{2-7}
\shortstack[lb]{Disciplines\\~} & $N_{c}$ & D & $N_{c}$ & D & $N_{c}$ & D \\
\cline{1-7}
Physics (1901-1921)& 2270.81&	50.27 & 37.24&	86.63 & 46.55&	68.47\\
\cline{1-7}
Chemistry (1901-1927)& 1159.67	&70.83 & 131.76&	84.02 & 9.50	&77.91\\
 \cline{1-7}
Physiology/Medicine (1901-1931)& 1979.43	&60.39 & 63.10	&81.70 & 19.30	&81.27\\
\hline 
\end{tabular}
\caption{Average citation count ($N_{c}$) per scientist (considering 21 scientists in each discipline) and diversity value (D) for total citation, total publication, and per-paper citation across three disciplines in past era.}
\label{tab5}
\end{table}
It is evident that, in earlier period, the diversity values for physics were the lowest in both the total citation and per-paper citation perspectives. However, in terms of the total publication perspective, physics had the highest diversity value. This suggests that citation diversity in physics was significant in previous times, whether considering the total citations or per-paper citations of award winners in this discipline. Conversely, the number of papers in physics was more evenly distributed compared to the other two subjects in historical time, as indicated by the higher diversity value for total publications. Additionally, the average citation count for chemistry was the lowest in both total citation and per-paper citation perspectives, while for the total publication, physics had the minimum average citation count. This again demonstrates that to obtain an exhaustive understanding of citation analysis, it is essential to look at the citation diversity values along with the citation counts of the publication data.

\subsubsection{Comparing citation diversity between recent and old times award winners in three principal disciplines}

\begin{figure}[h]
     \centering
     \begin{subfigure}[b]{0.25\textwidth}
         \centering
         \includegraphics[width=\textwidth]{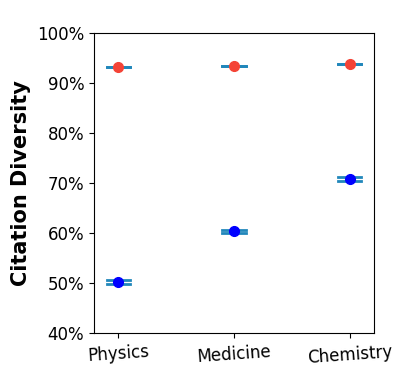}
         \caption{Total citation.}
         \label{fig:7a}
     \end{subfigure}
     \hfill
     \begin{subfigure}[b]{0.25\textwidth}
         \centering
         \includegraphics[width=\textwidth]{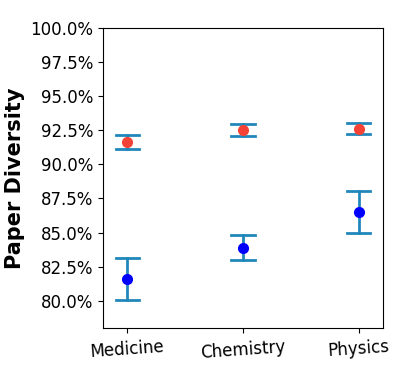}
         \caption{Total publication.}
         \label{fig:7b}
     \end{subfigure}
     \hfill
     \begin{subfigure}[b]{0.25\textwidth}
         \centering
         \includegraphics[width=\textwidth]{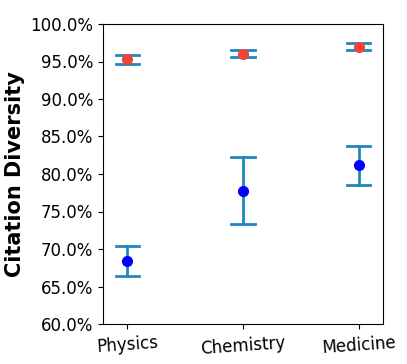}
         \caption{Per-paper citation.}
         \label{fig:7c}
     \end{subfigure}
     \caption{Visual representation of the comparison of  diversity (D) among award winning scientists across three principal disciplines for all three perspectives. Here red dots represent the diversity measure of each award winners in recent times and dark blue dots are the same for previous times, along with their 95\% confidence intervals (blue vertical lines), for the both cases.}
        \label{fig:comp}
\end{figure}

We now 
compare the citation diversity values 
across three principal disciplines between recent and old times. 
Table~\ref{tab5a} shows the D values for both recent and past times from 3 different viewpoints.   
Fig.~\ref{fig:comp} clearly illustrates a significant increase in D in recent times for all three disciplines in all cases. 
\begin{table}[h]
 \centering
 \small
\begin{tabular}{|c|c|c|c|c|c|c|}
\hline
 & \multicolumn{3}{c|}{D in Recent times} & \multicolumn{3}{c|}{D in Old times}\\
 \cline{2-7}
\shortstack[lb]{Disciplines\\~} & \shortstack[lb]{total\\citation} & \shortstack[lb]{total \\ publication} & \shortstack[lb]{per-paper\\ citation}& \shortstack[lb]{total\\citation} & \shortstack[lb]{total \\ publication} & \shortstack[lb]{per-paper\\ citation}\\
 \cline{1-7}
Physics &93.91 & 92.73& 95.41&50.27&86.63&68.47\\
\cline{1-7}
Chemistry & 93.38&92.65 &97.14&70.83&84.02&77.91\\
 \cline{1-7}
Physiology/Medicine &93.52 &91.77 &96.20&60.39&81.70&81.27\\
\hline 
\end{tabular}
\caption{Comparison of diversity measure (D) of three principal disciplines in recent and old times.}
\label{tab5a}
\end{table}

From this comparison, we can see that diversity values have increased across all disciplines from past eras to recent times. Notably, the recent era shows very small differences in diversity values among the three disciplines, highlighting a more even distribution of total number of citations, total number of papers, and per-paper citations among award winners across all disciplines. In contrast, these differences were much larger in the past. Additionally, we observe that while physics had the lowest diversity for the total citation in earlier times, it now has the highest D value compared to the other two disciplines in recent times, suggesting a significant improvement in the equality of citation distribution for physics. Again Fig.~\ref{fig:comp} reveals that the confidence interval for the total citation perspective is minimal, whereas the confidence intervals for the total publication and per-paper citation are comparatively large in both the recent and past eras. 
Thus, we can infer that the diversity estimate for the total citation is more reliable compared to the other two perspectives. Overall, this comparison reveals that the citation distribution in physics has improved markedly in recent times compared to previous times.

\subsection{Citation diversity in the publication of Individual prize winning scientists}
We now aspire to inspect the citation diversity in the publications of the individual prize winning scientists across various scientific disciplines. We  have chosen the citation data for a total of 135 eminent scholars from the aforesaid scientific disciplines. In particular, we have considered the Nobel prize winners in physics (30), chemistry (30), physiology/medicine (30) and economics (15), Abel prize winners in mathematics (15) and Turing award winners in computer science (15). 
\subsubsection{The Nobel Prize winners in Physics}
\begin{figure}[h]
     \centering
     \begin{subfigure}[b]{0.48\textwidth}
         \centering
         \includegraphics[width=\textwidth]{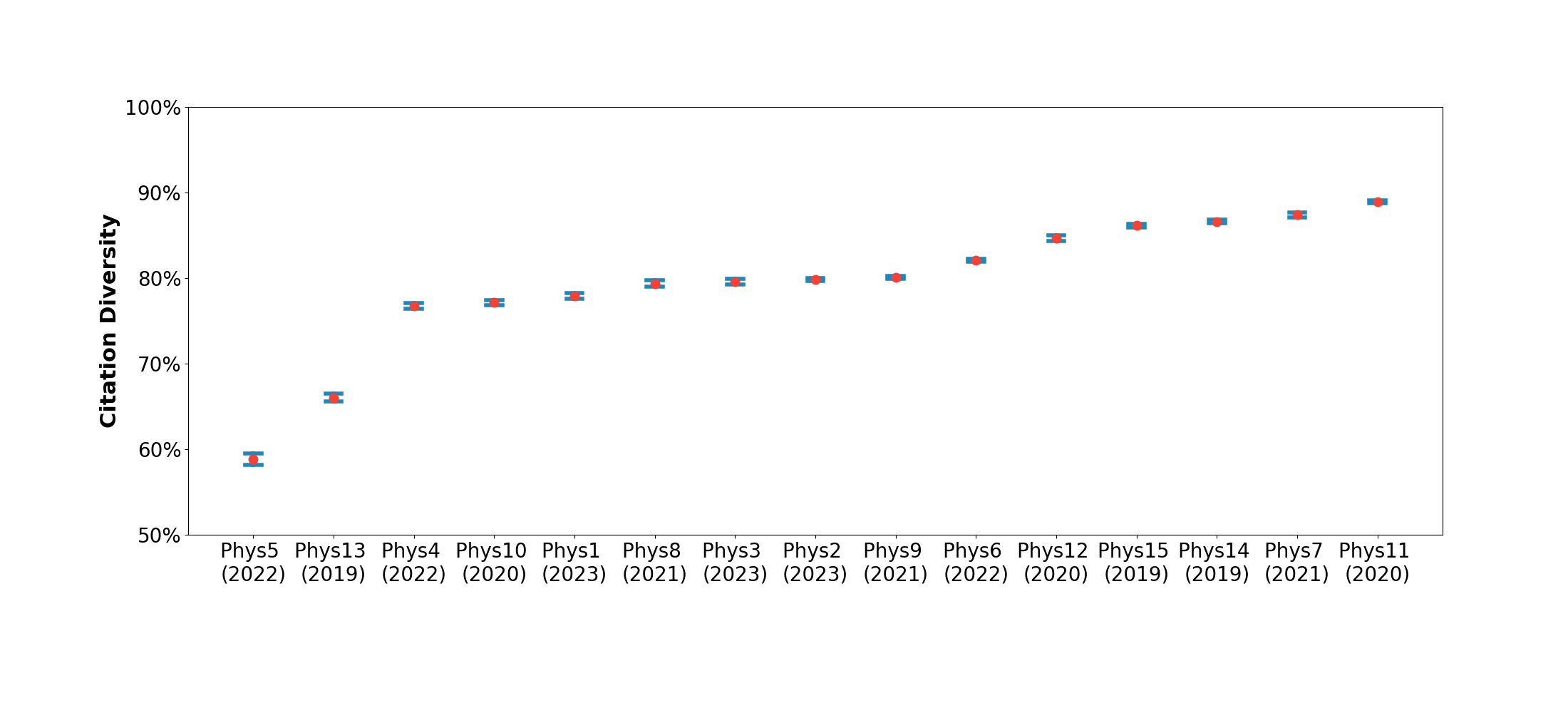}
         \caption{Citation diversity (D) with confidence intervals.}
         \label{fig:8a}
     \end{subfigure}
     \hfill
     \begin{subfigure}[b]{0.48\textwidth}
         \centering
         \includegraphics[width=\textwidth]{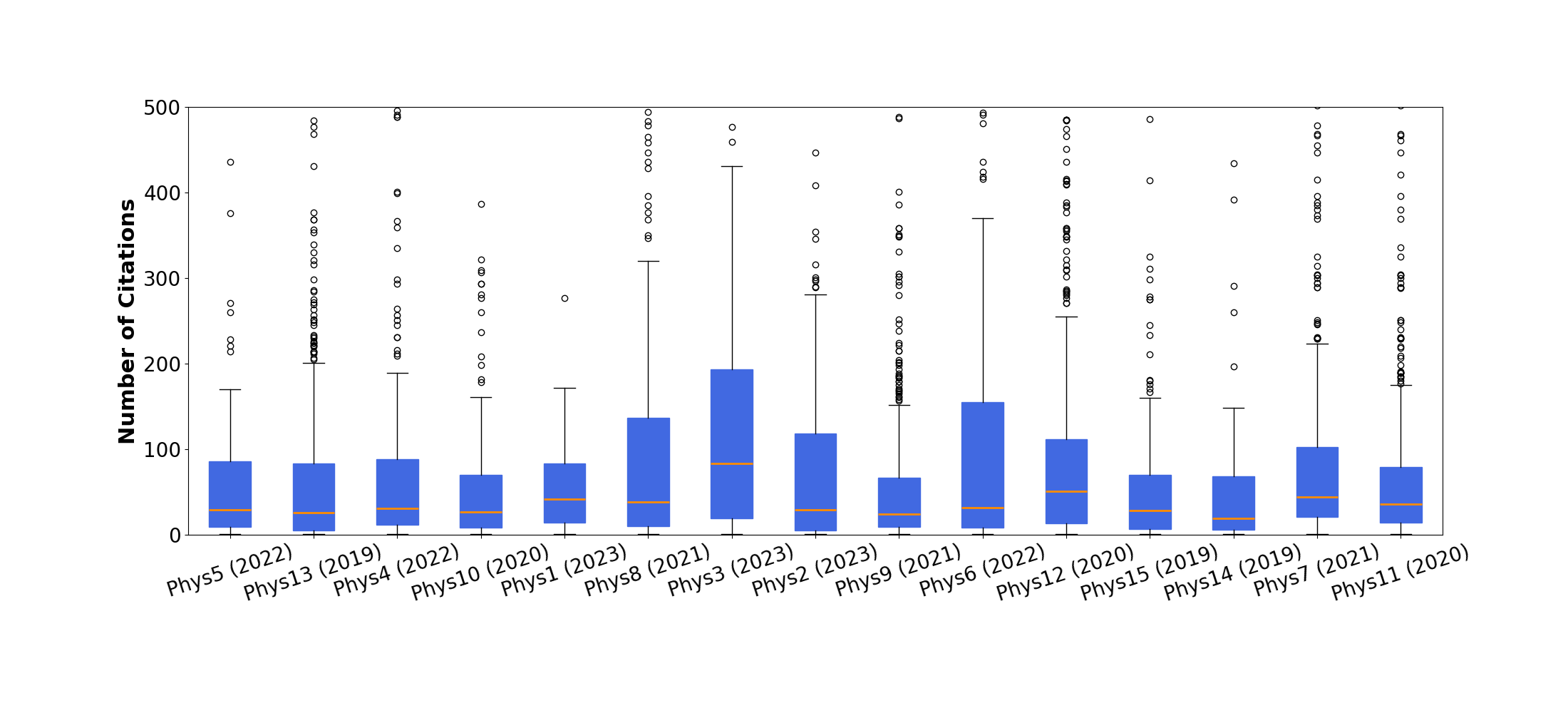}
         \caption{Box-plot of the total citation count.}
         \label{fig:8b}
     \end{subfigure}
        \caption{(a) Citation diversity (red dots) of recent Nobel laureates during (2019-2023) in physics, along with their 95\% confidence intervals (blue vertical lines) and (b) box-plot for the citation counts of each of them.}
        \label{fig:Physics_new}
\end{figure}

\begin{figure}[!h]
     \centering
     \begin{subfigure}[b]{0.48\textwidth}
         \centering
         \includegraphics[width=\textwidth]{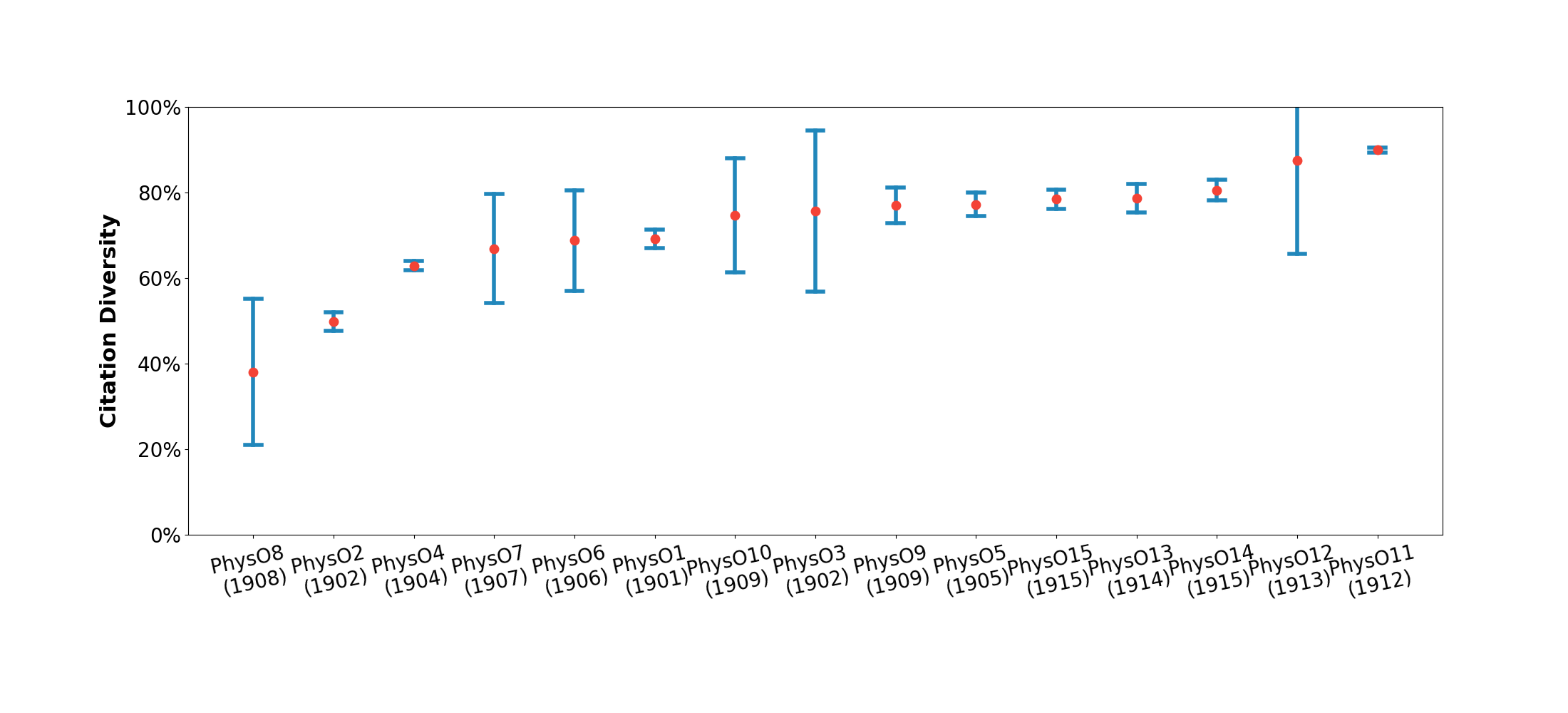}
         \caption{Citation diversity (D) with confidence intervals.}
         \label{fig:9a}
     \end{subfigure}
     \hfill
     \begin{subfigure}[b]{0.48\textwidth}
         \centering
         \includegraphics[width=\textwidth]{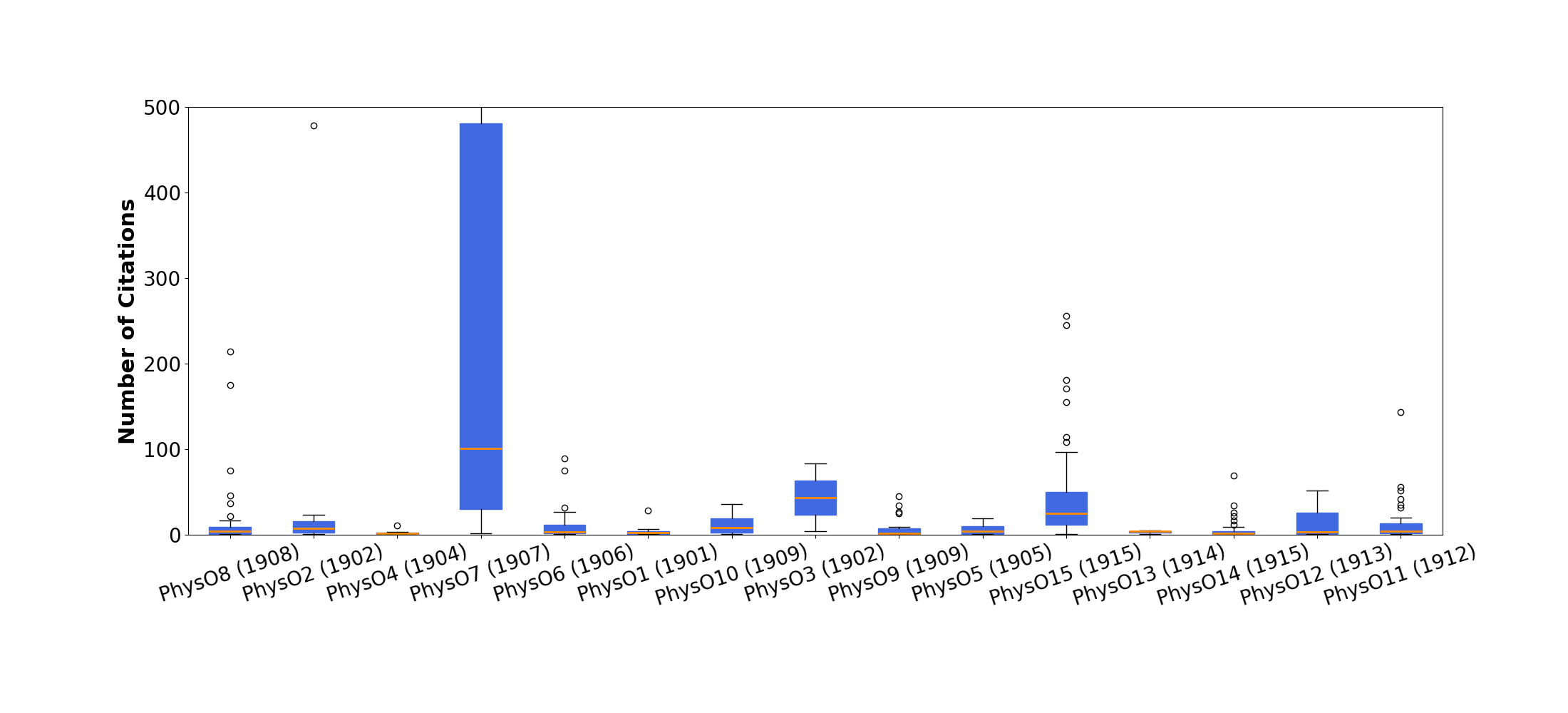}
         \caption{Box-plot of the total citation count.}
         \label{fig:9b}
     \end{subfigure}
        \caption{(a) Citation diversity (red dots) of earlier Nobel laureates during (1901-1915) in physics, along with their 95\% confidence intervals (blue vertical lines) and (b) box-plot for citation counts of each of them.}
        \label{fig:Physics_old}
\end{figure}

The Nobel prize in physics has been awarded to 224 individuals 
between 1901 and 2023. 
For our investigation we have explored  the citation counts 
of 30 Nobel laureates in physics, 15 from recent times (2019-2023) and 15 from the period (1901-1915). 
Using this data, we calculated the citation diversity values in the publication of these scientists 
following the methodology outlined above in Section III. 
Table~\ref{tab6} provides the citation diversity (D) values for each scientist considered along with their average citation counts ($N_{c}$). 
Fig.~\ref{fig:Physics_new} illustrates the citation diversity values for each recent laureate, with Fig.~\ref{fig:8a} specifically highlighting the citation diversity of recent laureates. Notably, the confidence intervals for each point are very small, confirming the accuracy of these values. Additionally, Fig.~\ref{fig:8b} presents a box-plot of the citation counts of the laureates from raw citation data of their publications. The citation diversity values for the 15 recent laureates range from about 60\% to 90\%, with higher diversity correlating with lower average citation counts, underscoring the limitations of using average citations alone to represent a laureate's citation distribution. 
In Fig.~\ref{fig:Physics_old}, we observe the citation diversity values of earlier Nobel laureates, with Fig.~\ref{fig:9a} revealing a wide range of citation diversity from 20\% to 80\%. Larger confidence intervals further extend this range. Fig.~\ref{fig:9b} shows a box-plot based on their raw citation data, revealing that earlier laureates generally had lower citation counts but higher diversity values.  Overall, the increase in estimated diversity values from earlier to more recent laureates suggests a decline in citation diversity among Nobel laureates in physics over the years.
\begin{table}[h]
 \centering
 \small
\begin{tabular}{|l|r|c|}
\hline
Nobel Laureates & $N_{c}$ & D \\
\hline
Phys1 (2023)  & 105.72	&77.83\\
\hline
Phys2 (2023)  & 112.46&	79.88\\
\hline
Phys3 (2023)  & 95.74	&79.49\\
\hline
Phys4 (2022)  & 102.42&	76.70\\
\hline
Phys5 (2022)   & 314.39	&58.92\\
\hline
Phys6 (2022)  & 150.22&	82.08\\
\hline
Phys7 (2021)   & 138.68	&87.46\\
\hline
Phys8 (2021)  & 128.70&	79.32\\
\hline
Phys9 (2021)  & 79.60&	80.02\\
\hline
Phys10 (2020)  & 171.86	&77.10\\
\hline
Phys11 (2020)  & 101.63	&88.94\\
\hline
Phys12 (2020)  & 66.37	&84.61\\
\hline
Phys13 (2019) & 129.35	&65.91\\
\hline
Phys14 (2019) & 97.12&	86.55\\
\hline
Phys15 (2019)  & 73.71	&85.07\\
\hline
\end{tabular}
 \begin{tabular}{|l|r|c|}
\hline
Nobel Laureates & $N_{c}$ & D \\
\hline
PhysO1 (1901)&	19.42&	69.21\\
\hline
PhysO2 (1902)&	86.19&	49.77\\
\hline
PhysO3 (1902)&	3.17&	75.90\\
\hline
PhysO4 (1904)&	477.40&	62.97\\
\hline
PhysO5 (1905)&	11.84&	77.17\\
\hline
PhysO6 (1906)&	5.75&	68.77\\
\hline
PhysO7 (1907)&	13.50&	67.09\\
\hline
PhysO8 (1908)&	43.50&	37.51\\
\hline
PhysO9 (1909)&	8.70&	77.11\\
\hline
PhysO10 (1909)&	7.00&	74.56\\
\hline
PhysO11 (1912)	&40.85&	90.01\\
\hline
PhysO12 (1913)&	3.33&	87.38\\
\hline
PhysO13 (1914)&	6.79&	78.61\\
\hline
PhysO14 (1915)&	13.58&	80.42\\
\hline
PhysO15 (1915)&	13.43&	78.39\\
\hline
\end{tabular}
\caption{Average citation count ($N_{c}$) per publication and citation diversity (D) of 15 Nobel laureates during (2019-2023) and 15 laureates during (1901-1915) in physics.}
\label{tab6}
\end{table}

\subsubsection{Nobel Prize winners in Chemistry} 
\begin{figure}[h]
     \centering
     \begin{subfigure}[b]{0.48\textwidth}
         \centering
         \includegraphics[width=\textwidth]{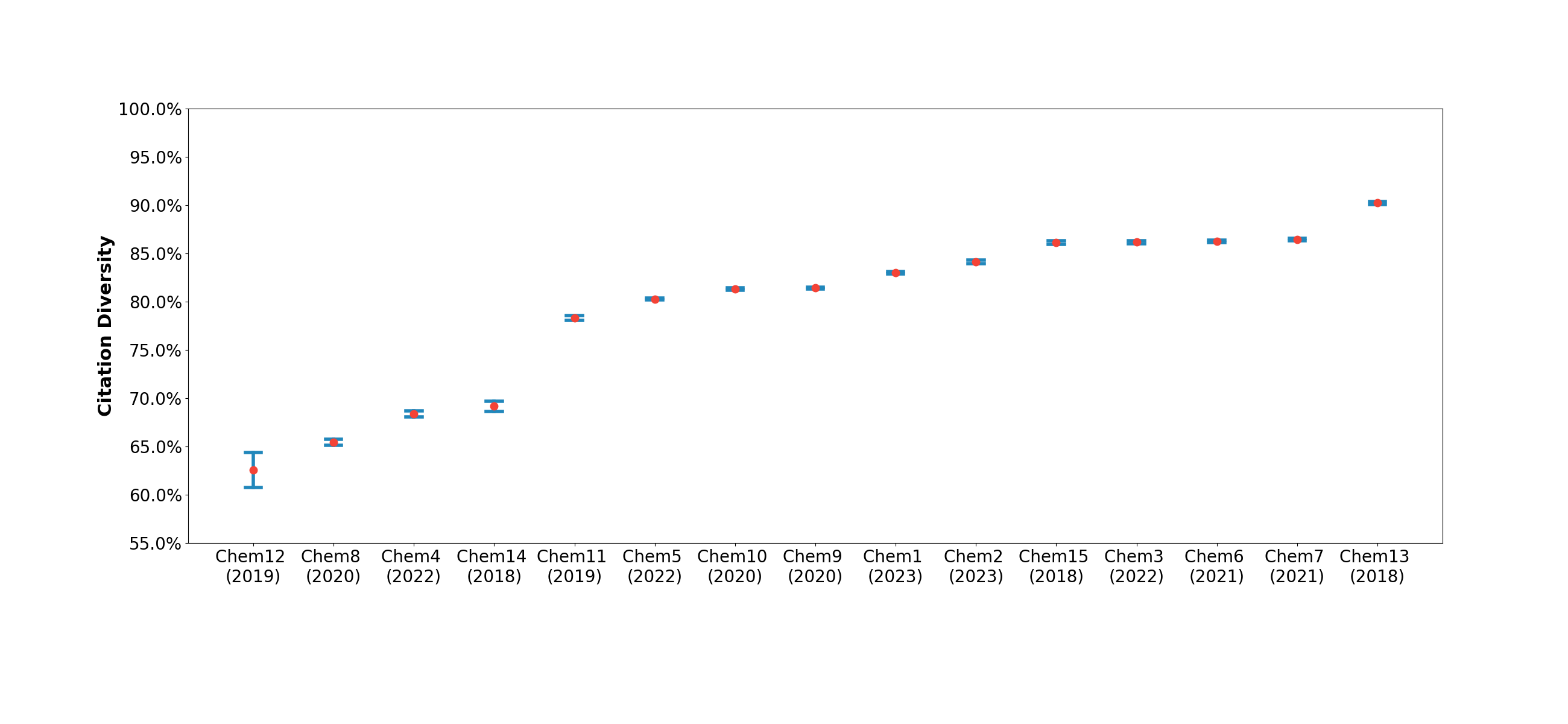}
         \caption{Citation diversity (D) with confidence intervals.}
         \label{fig:10a}
     \end{subfigure}
     \hfill
     \begin{subfigure}[b]{0.48\textwidth}
         \centering
         \includegraphics[width=\textwidth]{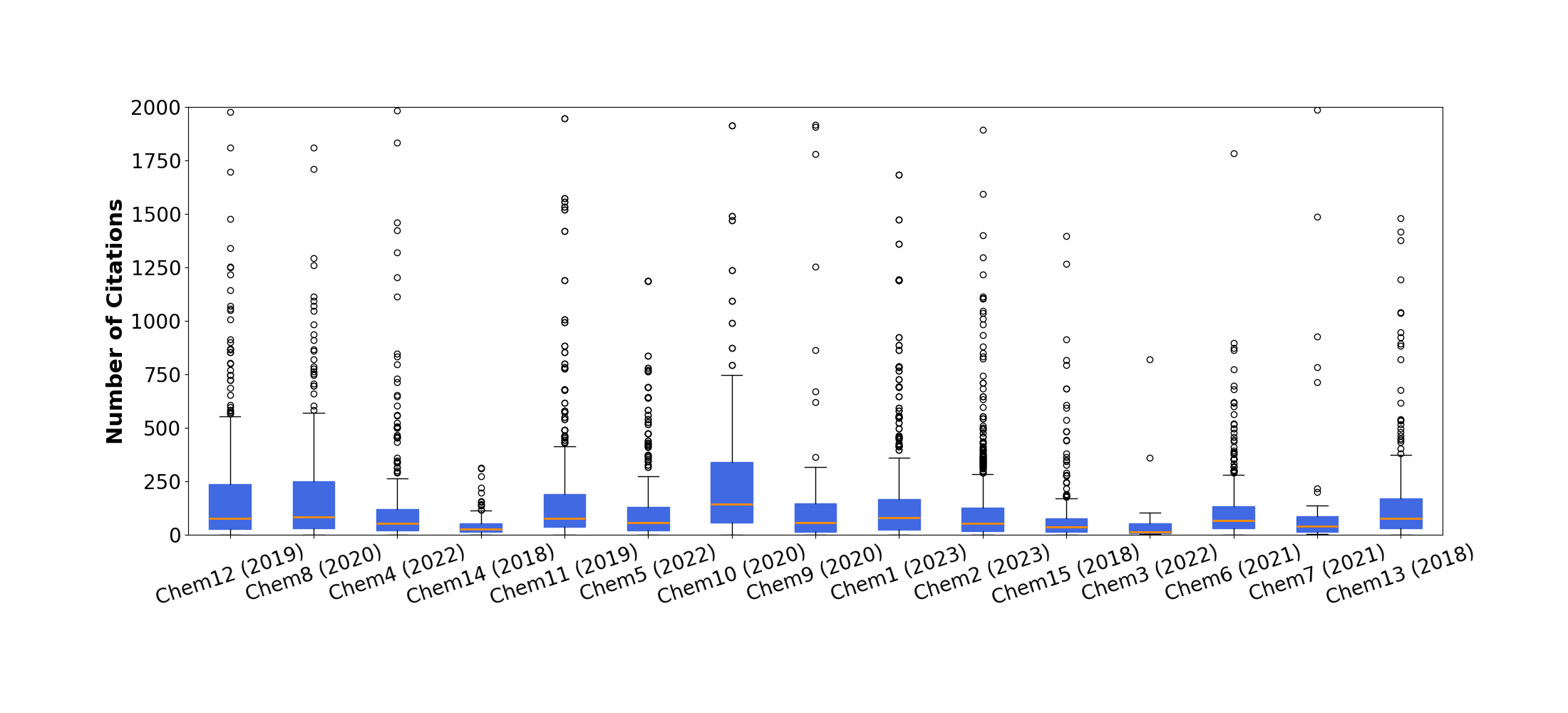}
         \caption{Box-plot of the total citation count.}
         \label{fig:10b}
     \end{subfigure}
        \caption{(a) Citation diversity (red dots) of recent Nobel laureates during (2018-2023) in chemistry, along with their 95\% confidence intervals (blue vertical lines) and (b) box-plot for citation counts of each of them.}
        \label{fig:chem_new}
\end{figure}

\begin{figure}[h]
     \centering
     \begin{subfigure}[b]{0.48\textwidth}
         \centering
         \includegraphics[width=\textwidth]{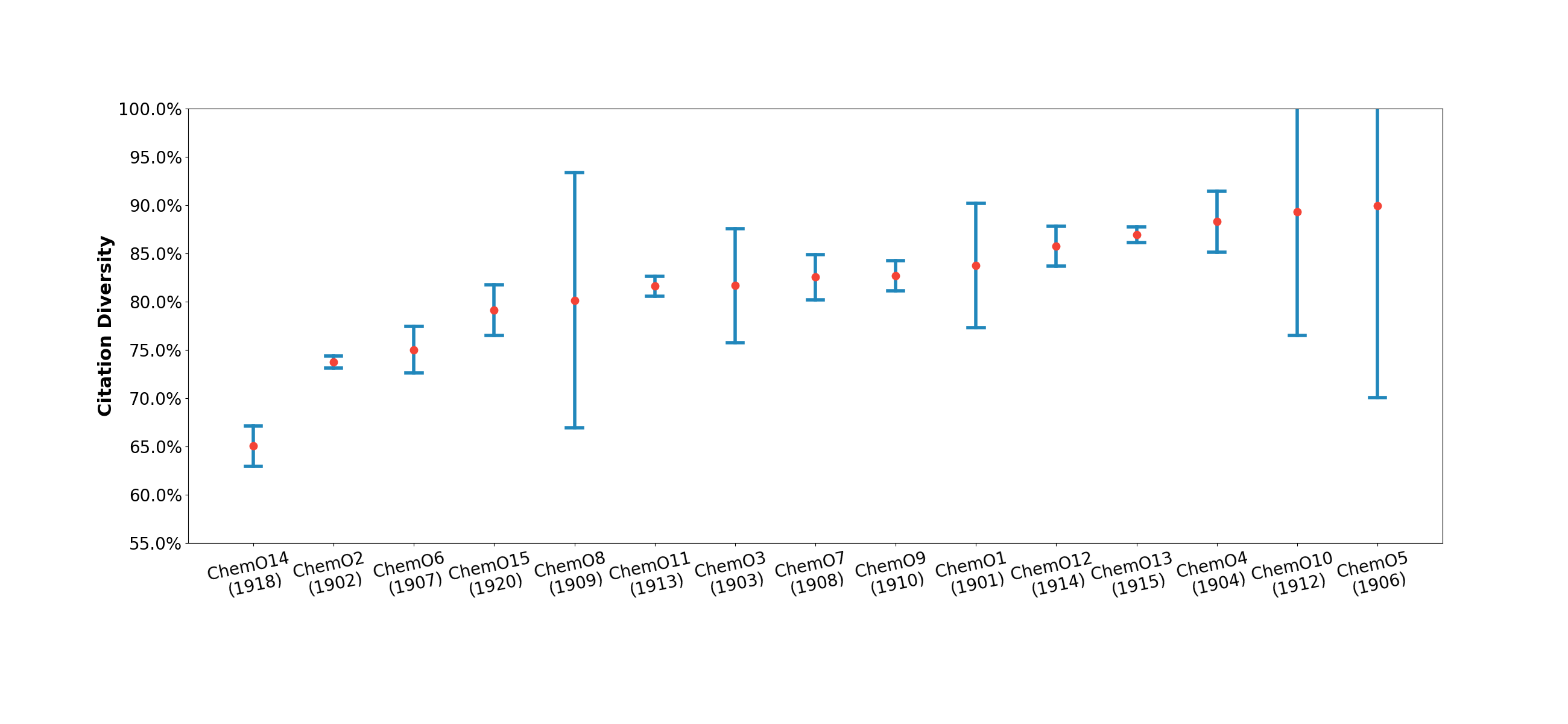}
         \caption{Citation diversity (D) with confidence intervals.}
         \label{fig:11a}
     \end{subfigure}
     \hfill
     \begin{subfigure}[b]{0.48\textwidth}
         \centering
         \includegraphics[width=\textwidth]{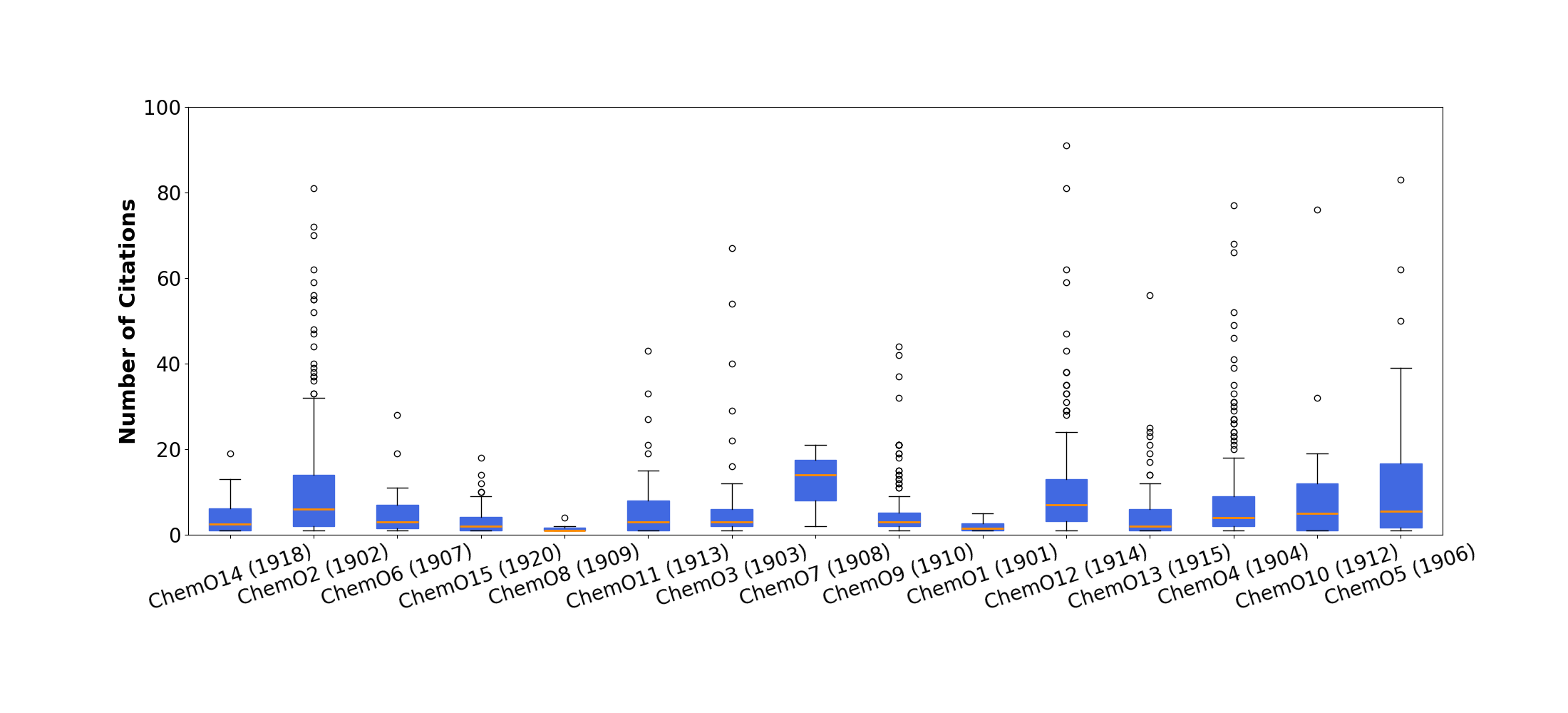}
         \caption{Box-plot of the total citation count.}
         \label{fig:11b}
     \end{subfigure}
        \caption{(a) Citation diversity (red dots) of old Nobel laureates during (1901-1920) in chemistry, along with their 95\% confidence intervals (blue vertical lines) and (b) box-plot for the citation counts of each of them.}
        \label{fig:Chem_old}
\end{figure}

From 1901 to 2023 the Nobel prize in chemistry has been bestowed on a total of 192 individuals.
For our analysis, we  have picked up the citation data of 30 Nobel prize winners in chemistry, 15 during 1901-1920 and 15 between  2018-2023.
Table~\ref{tab8} represents the calculated diversity values and average citation counts of all the listed Nobel prize winners in chemistry.
In Fig.~\ref{fig:chem_new}, we see the citation diversity of 15 recent Nobel laureates in chemistry, with diversity values ranging from 60\% to 90\%. Two distinct groups emerge: one containing four laureates with citation diversity between 60\%-70\%, indicating lower citation diversity, and another between 80\%-90\%, reflecting a more balanced citation distribution. The minimal confidence intervals confirm the reliability of these values. Fig.~\ref{fig:10b} further shows that despite similar average citation counts, diversity values differ significantly, revealing more insightful patterns in citation distribution. Meanwhile, Fig.~\ref{fig:Chem_old} shows earlier laureates' citation diversity ranging from 65\% to 90\%, though with larger confidence intervals, indicating greater variability. 
\begin{table}[h]
 \centering
 \small
 \begin{tabular}{|l|r|c|}
\hline
Nobel Laureates & $N_{c}$ & D\\
\hline
Chem1 (2023)&	252.39&	83.08\\
\hline
Chem2 (2023)&	245.78&	84.04\\
\hline
Chem3 (2022)	&123.02&86.09\\
\hline
Chem4 (2022)&	84.05&	68.28\\
\hline
Chem5 (2022)&	246.29	&80.23\\
\hline
Chem6 (2021)&	141.49	&86.24\\
\hline
Chem7 (2021)&	278.57&	86.46\\
\hline
Chem8 (2020)&	343.90&	65.34\\
\hline
Chem9 (2020)&	216.42&	81.47\\
\hline
Chem10 (2020)&	149.54	&81.26\\
\hline
Chem11 (2019)&	100.96&	78.22\\
\hline
Chem12 (2019)&	97.19&	62.76\\
\hline
Chem13 (2018)&	116.60&	90.24\\
\hline
Chem14 (2018)&	211.82&	69.12\\
\hline
Chem15 (2018)&	182.06&	86.16\\
\hline
\end{tabular}
\begin{tabular}{|l|r|c|}
\hline
Nobel Laureates & $N_{c}$ & D\\
\hline
ChemO1 (1901)&	5.00&	83.74\\
\hline
ChemO2 (1902)	&20.07&	73.72\\
\hline
ChemO3 (1903)&	6.20&	81.61\\
\hline
ChemO4 (1904)&	3.57&	88.26\\
\hline
ChemO5 (1906)&	1.67&	89.86\\
\hline
ChemO6 (1907)&	8.65&	75.06\\
\hline
ChemO7 (1908)&	7.38&	82.60\\
\hline
ChemO8 (1909)&	12.33&	80.10\\
\hline
ChemO9 (1910)&	6.52&	82.76\\
\hline
ChemO10 (1912)&	2.25&	89.45\\
\hline
ChemO11 (1913)&	15.81&	81.52\\
\hline
ChemO12 (1914)&	5.05&	85.77\\
\hline
ChemO13 (1915)&	9.00&	87.01\\
\hline
ChemO14 (1918)&	19.61&	65.01\\
\hline
ChemO15 (1920)	&15.75&	79.29\\
\hline 
\end{tabular}
\caption{Average citation count ($N_{c}$) per publication and citation diversity values (D) of 30 
Nobel laureates in chemistry, with 15 during the period (2018-2023) and 15 during the period (1901-1920).}
\label{tab8}
\end{table}
\subsubsection{Abel prize winners in Mathematics}
\begin{figure}[h]
     \centering
     \begin{subfigure}[b]{0.48\textwidth}
         \centering
         \includegraphics[width=\textwidth]{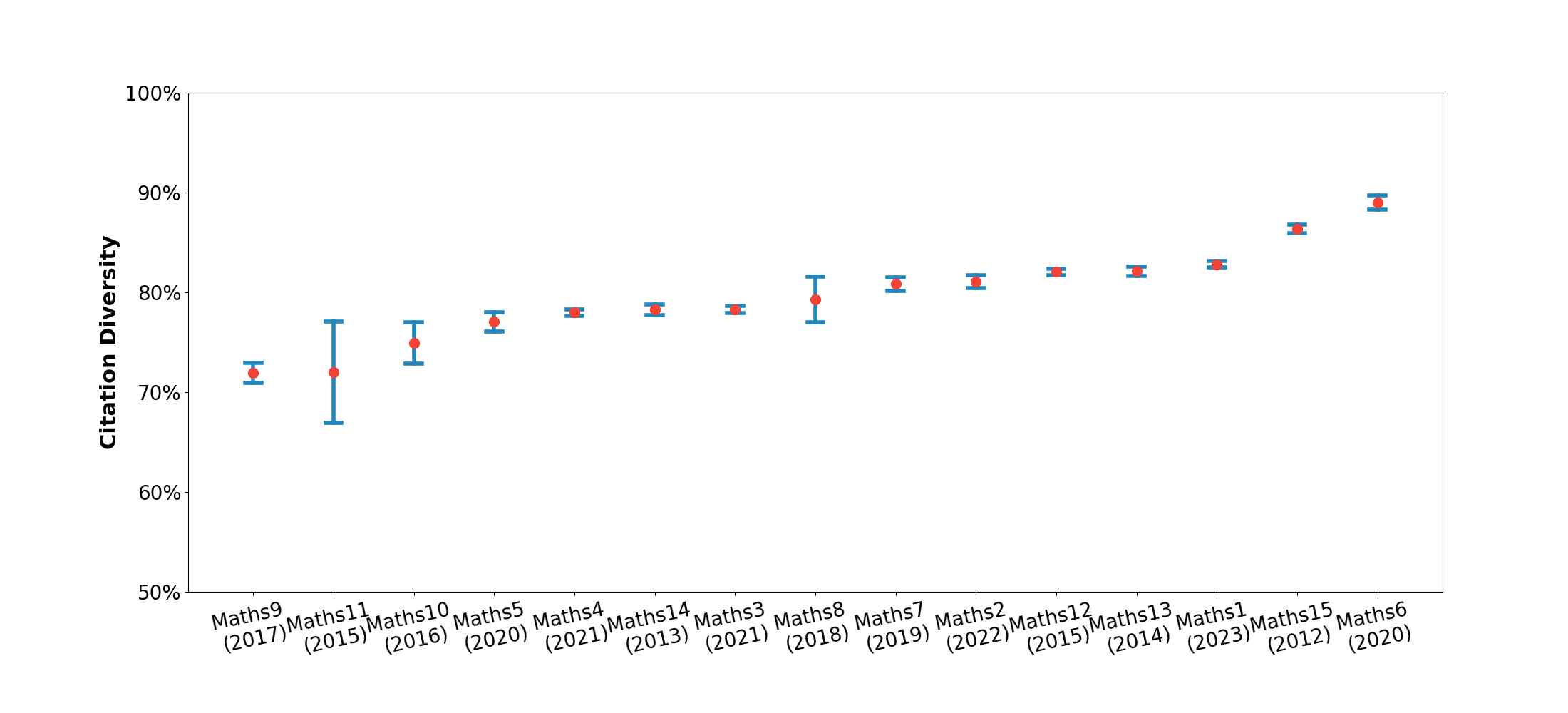}
         \caption{Citation diversity (D) with confidence intervals.}
         \label{fig:12a}
     \end{subfigure}
     \hfill
     \begin{subfigure}[b]{0.48\textwidth}
         \centering
         \includegraphics[width=\textwidth]{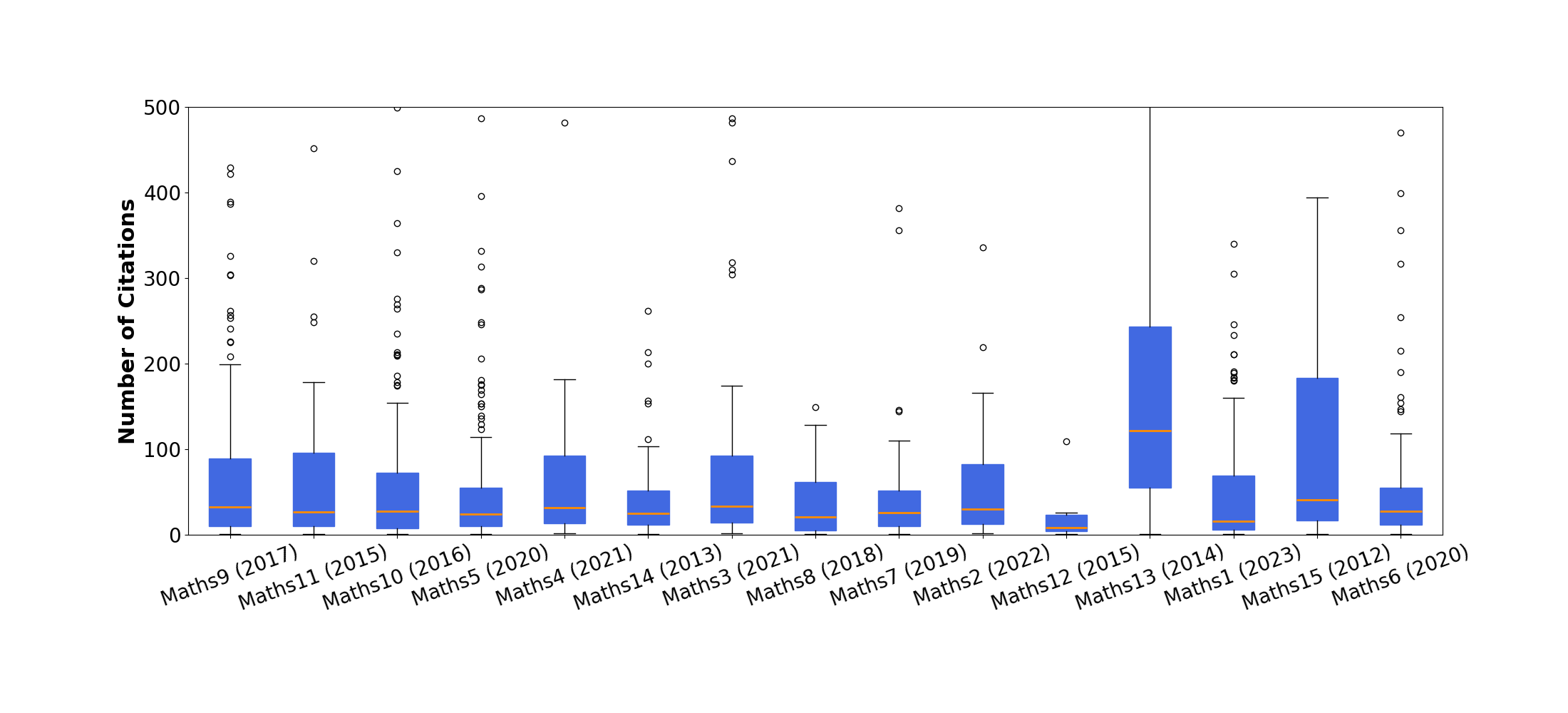}
         \caption{Box-plot of the total citation count.}
         \label{fig:12b}
     \end{subfigure}
        \caption{(a) Citation diversity (red dots) of Able prize winners during (2012-2023) in mathematics, along with their 95\% confidence intervals (blue vertical lines) and (b) box-plot for the citation counts of each of them.}
        \label{fig:Math}
\end{figure}

The Abel prize is awarded annually (2003 onwards) 
to one or more outstanding mathematicians 
and  is widely considered the Nobel prize of mathematics. 
Here we consider the data for 15 Able prize winners during 2012-2023. 
Table~\ref{tab10} presents the average citation counts and diversity values for each Abel prize winner considered.
In Fig.~\ref{fig:Math}, we observe the citation diversity and box-plot for the citation counts of each Abel prize winner. 
These citation diversity values of each scientist range from 70\% to 90\%, indicating moderate citation diversity. The confidence intervals for most diversity values are not high, though a few have slightly higher confidence intervals. This suggests that most calculated diversity values are reliable, with some variability due to the use of publicly available data. 

\begin{table}[h]
 \centering
 \small
\begin{tabular}{|l|r|c|}
\hline
Abel Prize Winners & $N_{c}$ & D\\
\hline
Maths1 (2023)&	86.14&	82.88\\
\hline
Maths2 (2022)&	81.47&	81.05\\
\hline
Maths3 (2021)&	84.41	&78.37\\
\hline
Maths4 (2021)&	71.79&	77.98\\
\hline
Maths5 (2020)&	99.16&	77.04\\
\hline
Maths6 (2020)&	43.90&	88.97\\
\hline
Maths7 (2019)&	109.15&	80.80\\
\hline
Maths8 (2018)&	43.69&	79.14\\
\hline
\end{tabular}
 \begin{tabular}{|l|r|c|}
\hline
Abel Prize Winners & $N_{c}$ & D\\
\hline
Maths9 (2017)&	75.45	&71.88\\
\hline
Maths10 (2016)&	76.58&	74.85\\
\hline
Maths11 (2015)&	21.20&	72.09\\
\hline
Maths12 (2015)&	289.14&	82.17\\
\hline
Maths13 (2014)&	58.38&	82.20\\
\hline
Maths14 (2013)&	187.70&	78.34\\
\hline
Maths15 (2012)&	56.92&	86.29\\
\hline
\multicolumn{3}{c}{}\\
\end{tabular}
\caption{Average citation count ($N_{c}$) per publication and citation diversity (D) of 15 Abel prize winners during 2012 to 2023 in mathematics.}
\label{tab10}
\end{table}
In this case, we observe that one of the 2015 Abel Prize winners has the highest $N_{c}$ value but a lower D value. Conversely, one of the 2020 Abel Prize winners has a comparatively lower $N_{c}$ but the highest D value. It is also noteworthy that, despite having similar $N_{c}$ values (with the exception of three cases), their D values vary significantly, ranging from 71\% to 89\%.
\subsubsection{Turing Award winners in Computer Science}
\begin{figure}[h]
     \centering
     \begin{subfigure}[b]{0.48\textwidth}
         \centering
         \includegraphics[width=\textwidth]{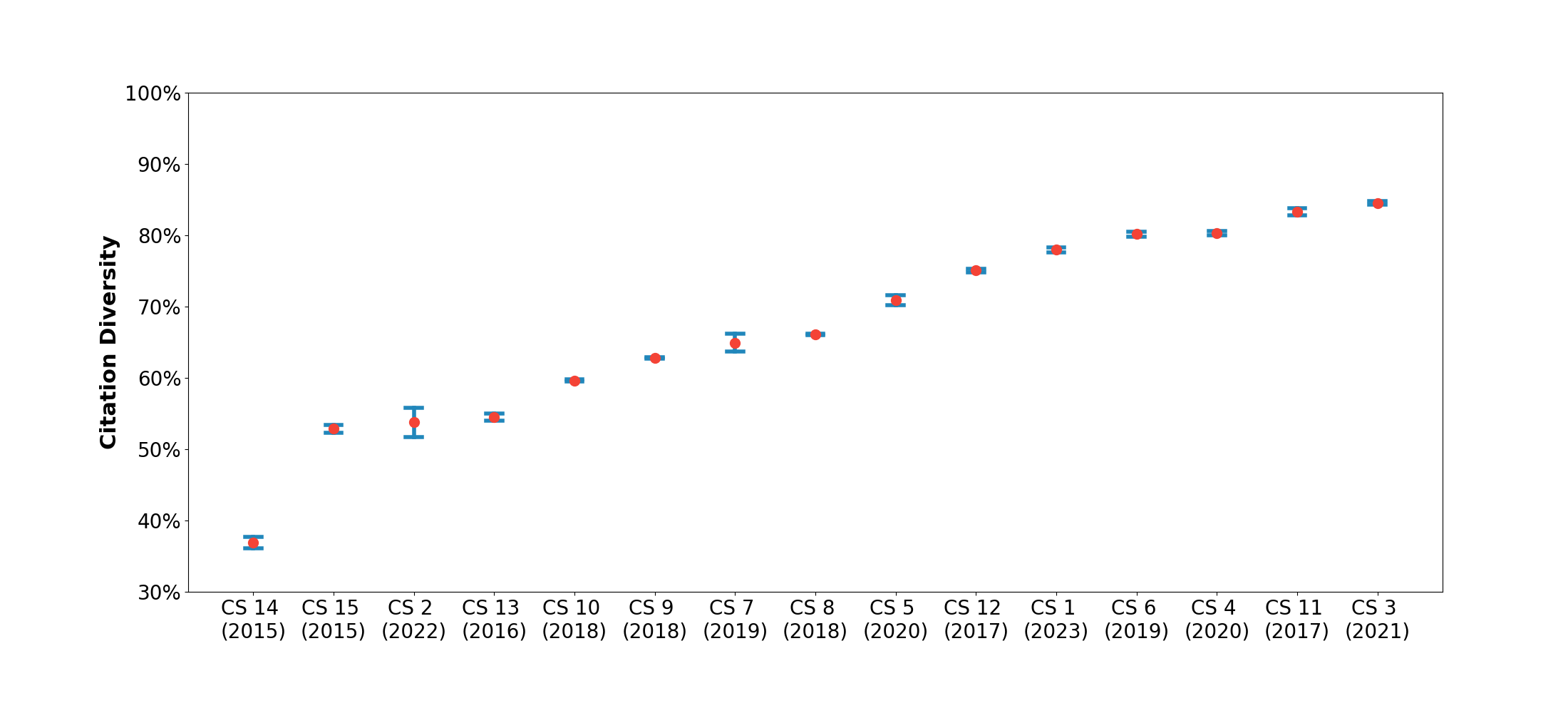}
         \caption{Citation diversity (D) with confidence intervals.}
         \label{fig:13a}
     \end{subfigure}
     \hfill
     \begin{subfigure}[b]{0.48\textwidth}
         \centering
         \includegraphics[width=\textwidth]{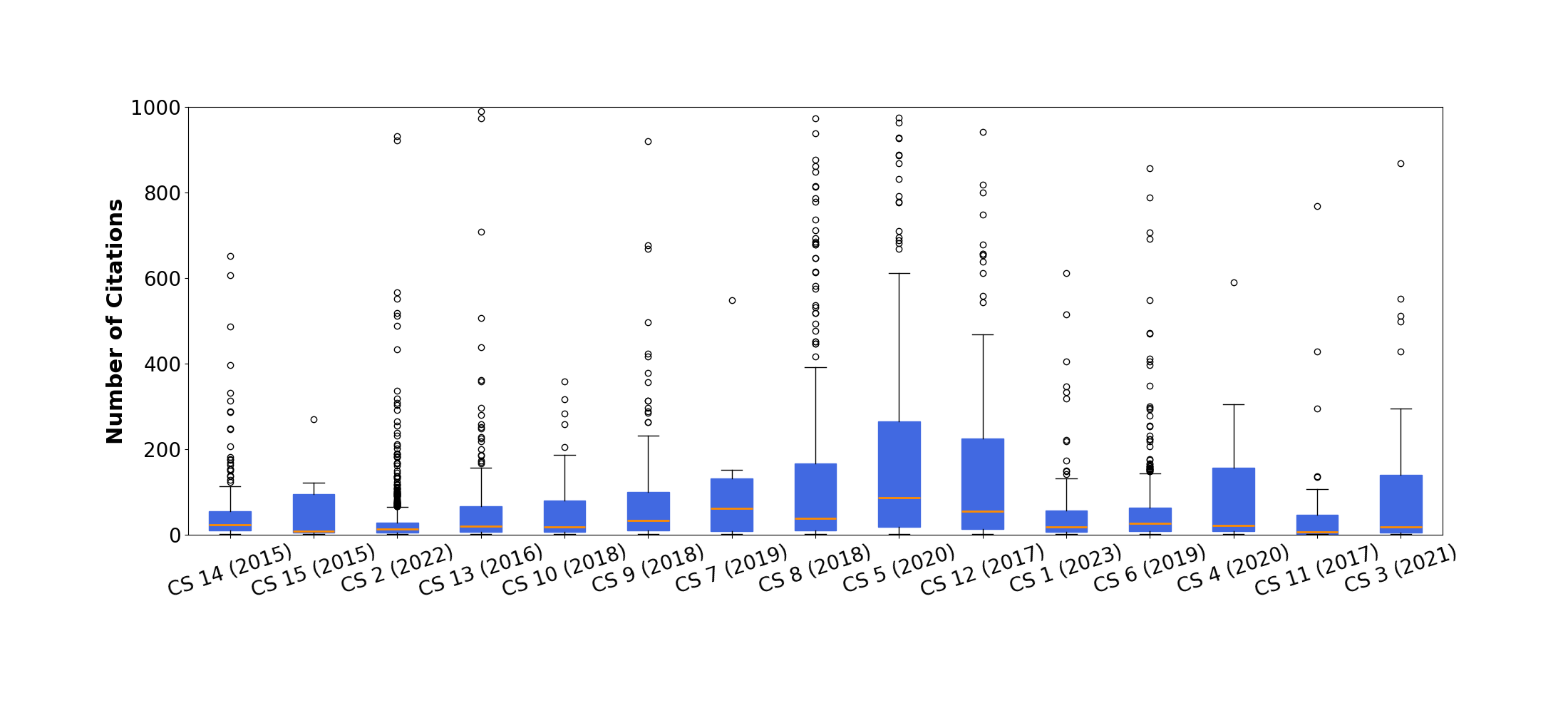}
         \caption{Box-plot of the total citation count.}
         \label{fig:13b}
     \end{subfigure}
        \caption{(a) Citation diversity (red dots) of Turing prize winners during (2015-2023) in computer science, along with their 95\% confidence intervals (blue vertical lines) and (b) box-plot for the citation counts of each of them.}
        \label{fig:Computer}
\end{figure}

The ACM A. M. Turing Award is an annual prize given by the Association for Computing Machinery (ACM) for contributions of lasting and major technical importance to computer science. 
Commencing in 1966, as of 2024, 77 people have been awarded the prize. We settled on the publication data of 15 highly recognized computer scientists between 2015-2023 only. 
Table~\ref{tab11} presents the calculated diversity values along with the average citation counts for each Turing prize winner.
\begin{table}[h]
 \centering
 \small
\begin{tabular}{|l|r|c|}
\hline
Turing Prize Winners & $N_{c}$ & D\\
\hline
CS 1 (2023)&	71.79&	77.98\\
\hline
CS 2 (2022)&	139.33&	53.93\\
\hline
CS 3 (2021)&	34.93&	84.51\\
\hline
CS 4 (2020)&	80.23&	80.22\\
\hline
CS 5 (2020)&	81.14&	70.10\\
\hline
CS 6 (2019)	&92.73&	80.18\\
\hline
CS 7 (2019)&	174.11	&65.04\\
\hline
CS 8 (2018)&	603.73&	65.10\\
\hline
\end{tabular}
 \begin{tabular}{|l|r|c|}
\hline
Turing Prize Winners & $N_{c}$ & D\\
\hline
CS 9 (2018)&	1365.96&	62.76\\
\hline
CS 10 (2018)&	683.66&	59.70\\
\hline
CS 11 (2017)&	54.85&	83.33\\
\hline
CS 12 (2017)&	88.96&	75.01\\
\hline
CS 13 (2016)	&362.10	&54.41\\
\hline
CS 14 (2015)&	411.17&	36.75\\
\hline
CS 15 (2015)&	349.94	&52.96\\
\hline
\multicolumn{3}{c}{}\\
\end{tabular}
\caption{Average citation count ($N_{c}$) per publication and citation diversity (D) of 15 Turing Prize winners during 2015 to 2023 in computer science.}
\label{tab11}
\end{table}
In Fig.~\ref{fig:Computer}, we present an 
overview of the diversity values and citation ranges for each Turing award winner in computer science. scientists
Notably, the diversity value of one of the 2015 Turing award winners is significantly lower compared to the others, indicating an uneven citation distribution for that individual scientist. Conversely, the diversity values of other computer scientists  range between 55\% and 85\%, with a distinct division at 70\%. Below this threshold, there are 7 Turing award winners, and above it, there are also 7 winners. This suggests that the citation diversity is lower for the 7 scientists (below 70\%) compared to those above it. In this analysis, the confidence interval is minimal, except for two cases. 
For better understanding, we maintain the same order of the award winners on the x-axis in Fig.~\ref{fig:13b} as in Fig.~\ref{fig:13a}, where they are arranged in ascending order of diversity values. However, no meaningful pattern is observed from the total citation range in Fig.~\ref{fig:13b} but there is a clear pattern in diversity values of each scientists shown in Fig.~\ref{fig:13a}.
\subsubsection{Nobel prize winners in Economics}
\begin{figure}[h]
     \centering
     \begin{subfigure}[b]{0.48\textwidth}
         \centering
         \includegraphics[width=\textwidth]{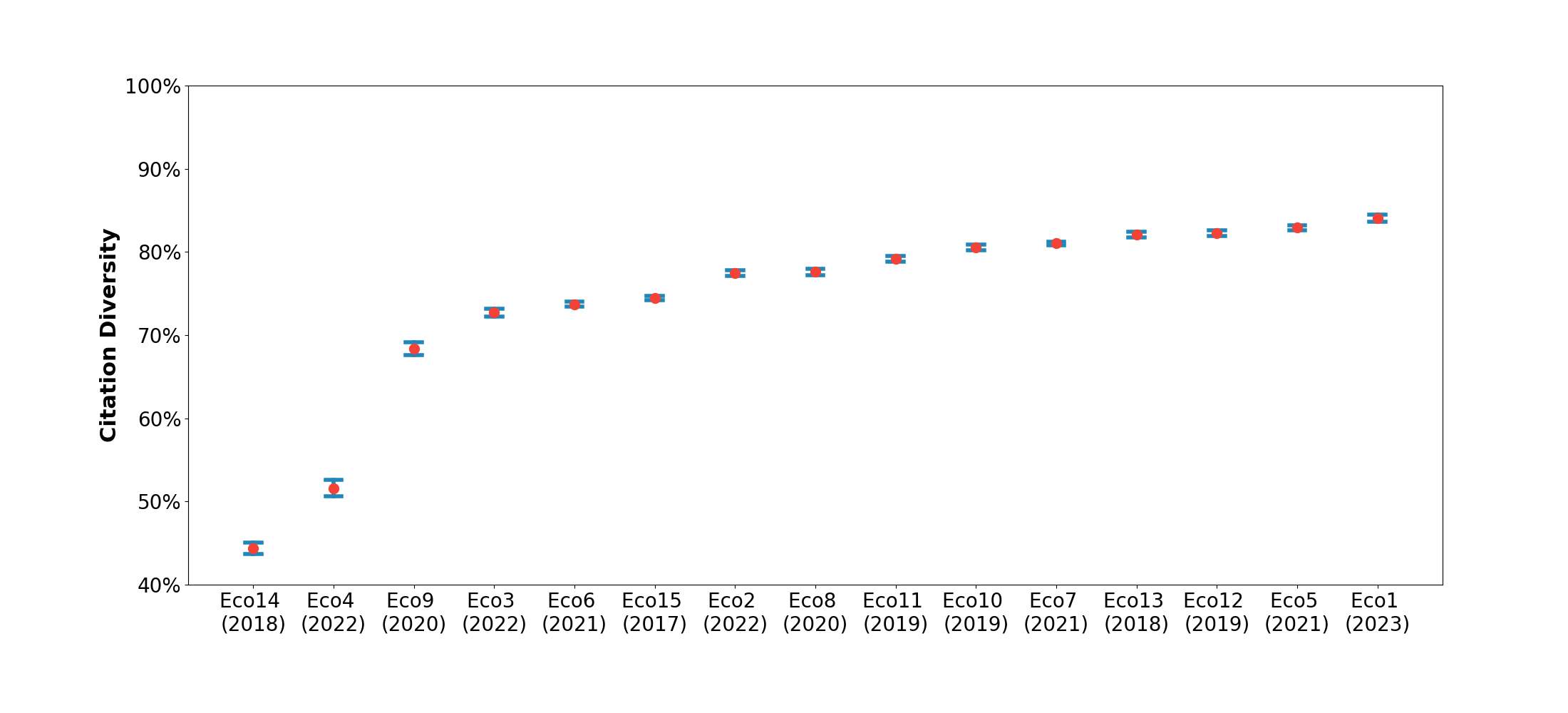}
         \caption{Citation diversity (D) with confidence intervals.}
         \label{fig:14a}
     \end{subfigure}
     \hfill
     \begin{subfigure}[b]{0.48\textwidth}
         \centering
         \includegraphics[width=\textwidth]{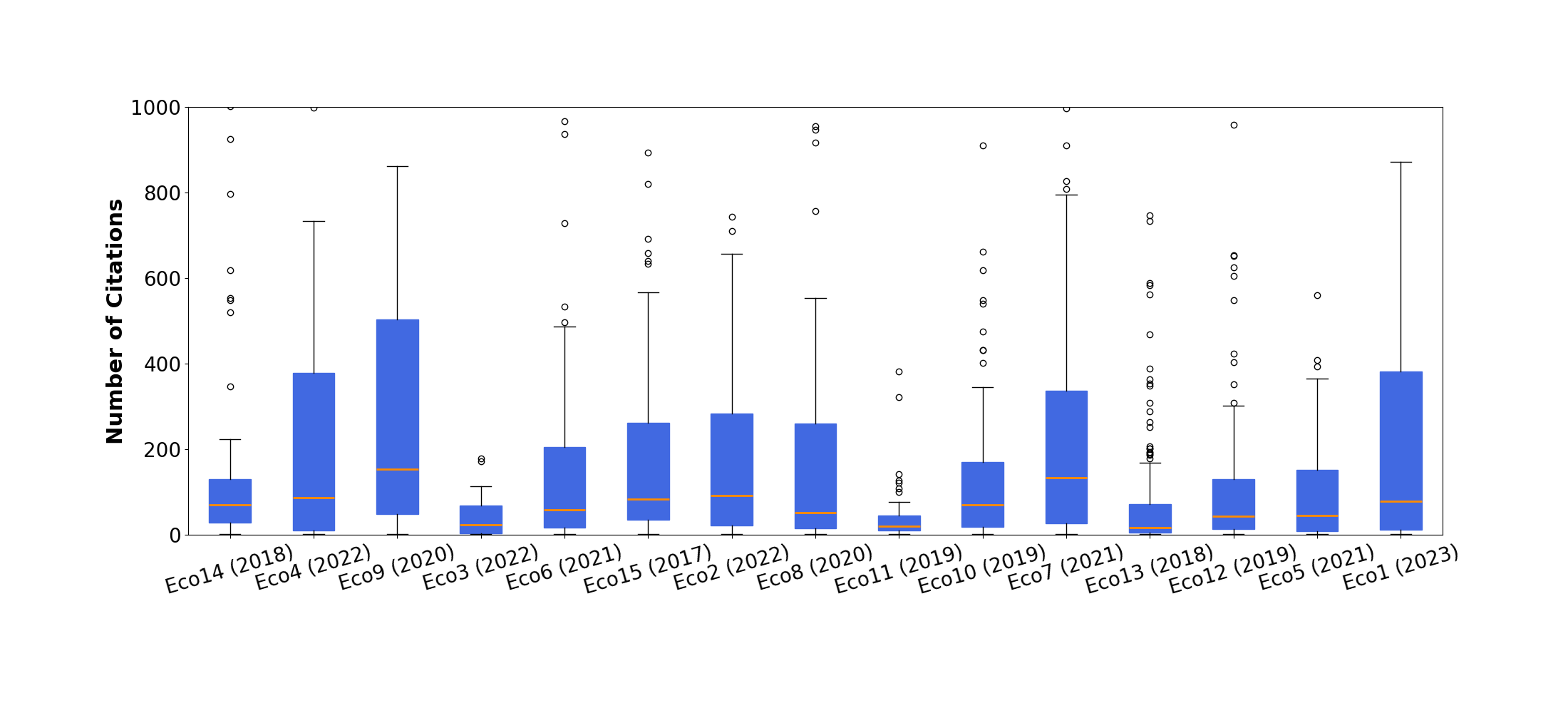}
         \caption{Box-plot of the total citation count.}
         \label{fig:14b}
     \end{subfigure}
        \caption{(a) Citation diversity (red dots) of Nobel laureates during (2017-2023) in economics, along with their 95\% confidence intervals (blue vertical lines) and (b) box-plot for the citation counts of each of them.}
        \label{fig:Economics}
\end{figure}

The first Nobel Memorial Prize in economic sciences  was awarded in 1969. 
As of 2023, 55 prizes in economic sciences have been given to 93 individuals. We considered the publication data of 15 distinguished economists between 2017-2023. 
Table \ref{tab12} presents the diversity values and average citation counts for these 15 individual Nobel laureates in economics. 
\begin{table}[h]
 \centering
 \small
 \begin{tabular}{|l|r|c|}
\hline
Nobel Laureates & $N_{c}$ & D\\
\hline
Eco1 (2023)	&147.54&	83.92\\
\hline
Eco2 (2022)&	352.53&	77.60\\
\hline
Eco3 (2022)&	554.38&	72.74\\
\hline
Eco4 (2022)&	143.22&	51.51\\
\hline
Eco5 (2021)&	166.18&	82.97\\
\hline
Eco6 (2021)&	299.72&	73.64\\
\hline
Eco7 (2021)&	313.47&	81.01\\
\hline
Eco8 (2020)&	259.66&	77.51\\
\hline
\end{tabular}
\begin{tabular}{|l|r|c|}
\hline
Nobel Laureates & $N_{c}$ & D\\
\hline
Eco9 (2020)&	95.00&	68.37\\
\hline
Eco10 (2019)&	169.13&	80.57\\
\hline
Eco11 (2019)&	279.92&	79.09\\
\hline
Eco12 (2019)&	72.81&	82.32\\
\hline
Eco13 (2018)&	143.30&	82.16\\
\hline
Eco14 (2018)&	521.79&	44.42\\
\hline
Eco15 (2017)&	402.01&	74.54\\
\hline
\multicolumn{3}{c}{}\\
\end{tabular}
\caption{Average citation count ($N_{c}$) per publication and citation diversity (D) of 15 Nobel prize winners in economics (2017-2023).}
\label{tab12}
\end{table}
Fig.~\ref{fig:14a} displays the citation diversity values for each laureate, showing that two recent laureates have very low diversity. In contrast, citation diversity of the other laureates 
ranging from 70\% to 85\%, implying 
a more balanced citation distribution across their publications. The small confidence intervals for these values reinforce the accuracy of our diversity calculations. Fig.~\ref{fig:14b} presents a box-plot of the total citations for each laureate, maintaining the same order of laureates as in Fig.~\ref{fig:14a} for easy comparison. Although the total citation counts vary among the laureates, it does not provide any significant insights into citation distribution. Instead, our diversity values effectively illustrate the extent of citation counts among these economics laureates.

\subsubsection{The Nobel Prize winners in Physiology/Medicine}
The Nobel prize in physiology/medicine has been awarded 114 times to a total of 227 laureates from 1901 to 2023. Our analysis focuses on two distinct groups: 15 laureates from the period 2017 to 2023 and another 15 from the period 1902 to 1923. 
Table \ref{tab13} presents the citation diversity values and corresponding average citation counts for all of their publications. 

\begin{table}[h]
 \centering
 \small
 \begin{tabular}{|l|r|c|}
\hline
Nobel Laureates & $N_{c}$ & D\\
\hline
Med1 (2023)&	111.00	&81.76\\
\hline
Med2 (2023)&	160.58&	81.65\\
\hline
Med3 (2022)&	194.66&	87.18\\
\hline
Med4 (2021)	&272.42	&87.25\\
\hline
Med5 (2021)&	405.67	&80.78\\
\hline
Med6 (2020)&	152.00&	88.11\\
\hline
Med7 (2020)	&117.09&	83.78\\
\hline
Med8 (2019)&	220.28&	86.70\\
\hline
Med9 (2019)&	287.89&	85.89\\
\hline
Med10 (2019)&	182.39&	82.40\\
\hline
Med11 (2018)&	239.69	&86.06\\
\hline
Med12 (2018)&	134.80	&84.59\\
\hline
Med13 (2017)&	109.78&	89.56\\
\hline
Med14 (2017)&	106.91&	91.27\\
\hline
Med15 (2017)&	123.39&	90.17\\
\hline
\end{tabular}
\begin{tabular}{|l|r|c|}
\hline
Nobel Laureates & $N_{c}$ & D\\
\hline
MedO1 (1902)&	11.61&	74.00\\
\hline
MedO2 (1903)&	8.00&	85.56\\
\hline
MedO3 (1904)&	21.75&	49.05\\
\hline
MedO4 (1905)&	12.08&	79.20\\
\hline
MedO5 (1906)&	42.05&	85.84\\
\hline
MedO6 (1908)&	12.42&	85.08\\
\hline
MedO7 (1910)&	58.13&	82.72\\
\hline
MedO8 (1911)&	3.67&	81.92\\
\hline
MedO9 (1914)&	9.24&	65.30\\
\hline
MedO10 (1920)&	59.63&	74.14\\
\hline
MedO11 (1922)&	9.72&	76.17\\
\hline
MedO12 (1923)&	66.69&	64.33\\
\hline
MedO13 (1923)&	2.82&	87.07\\
\hline
MedO14 (1926)&	7.43&	87.36\\
\hline
MedO15 (1927)	&20.00&	82.23\\
\hline
\end{tabular}
\caption{Average citation count ($N_{c}$) per publication and citation diversity (D) of 15 Nobel laureates during 2017 to 2023 and 15 laureates during 1902 to 1927 in physiology/medicine.}
\label{tab13}
\end{table}

\begin{figure}[h]
     \centering
     \begin{subfigure}[b]{0.48\textwidth}
         \centering
         \includegraphics[width=\textwidth]{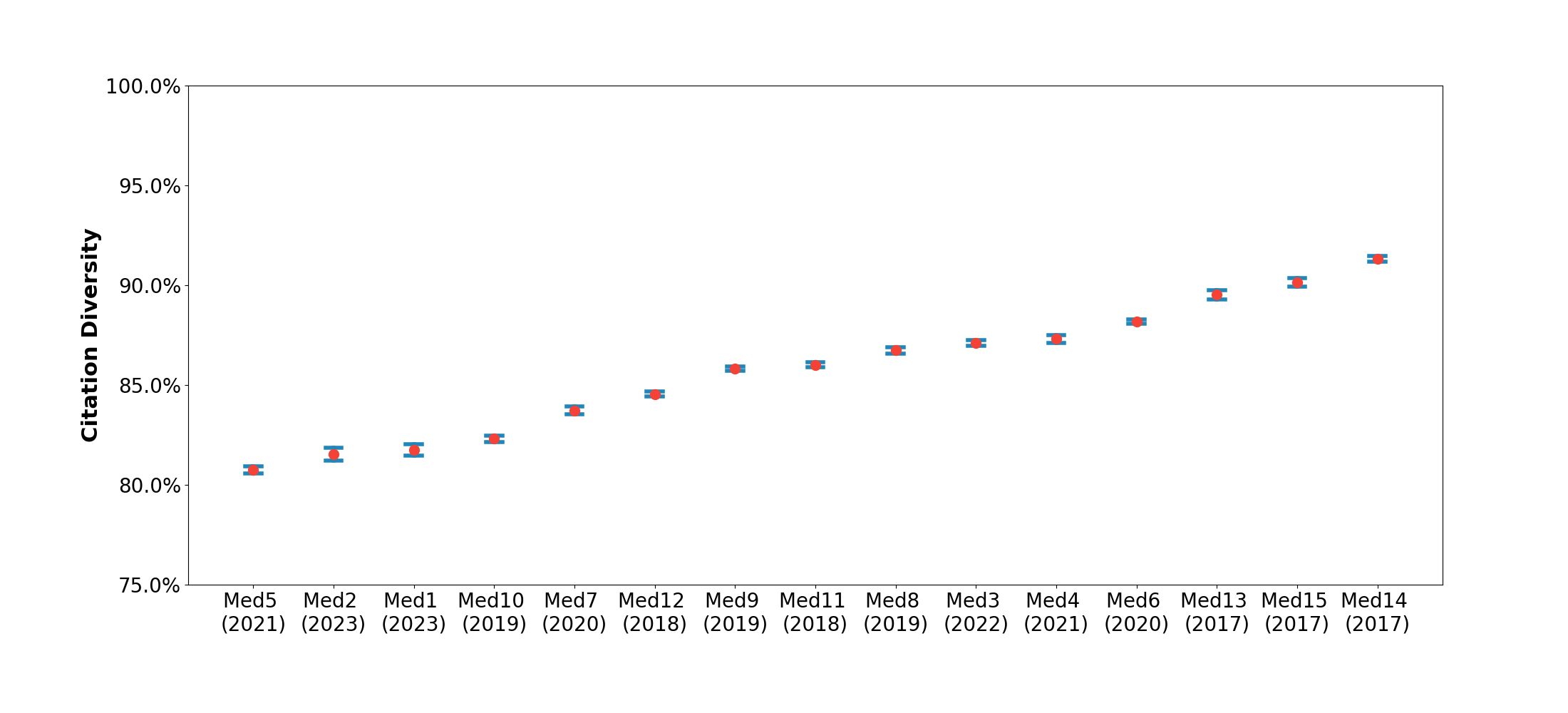}
         \caption{Citation diversity (D) with confidence intervals.}
         \label{fig:15a}
     \end{subfigure}
     \hfill
     \begin{subfigure}[b]{0.48\textwidth}
         \centering
         \includegraphics[width=\textwidth]{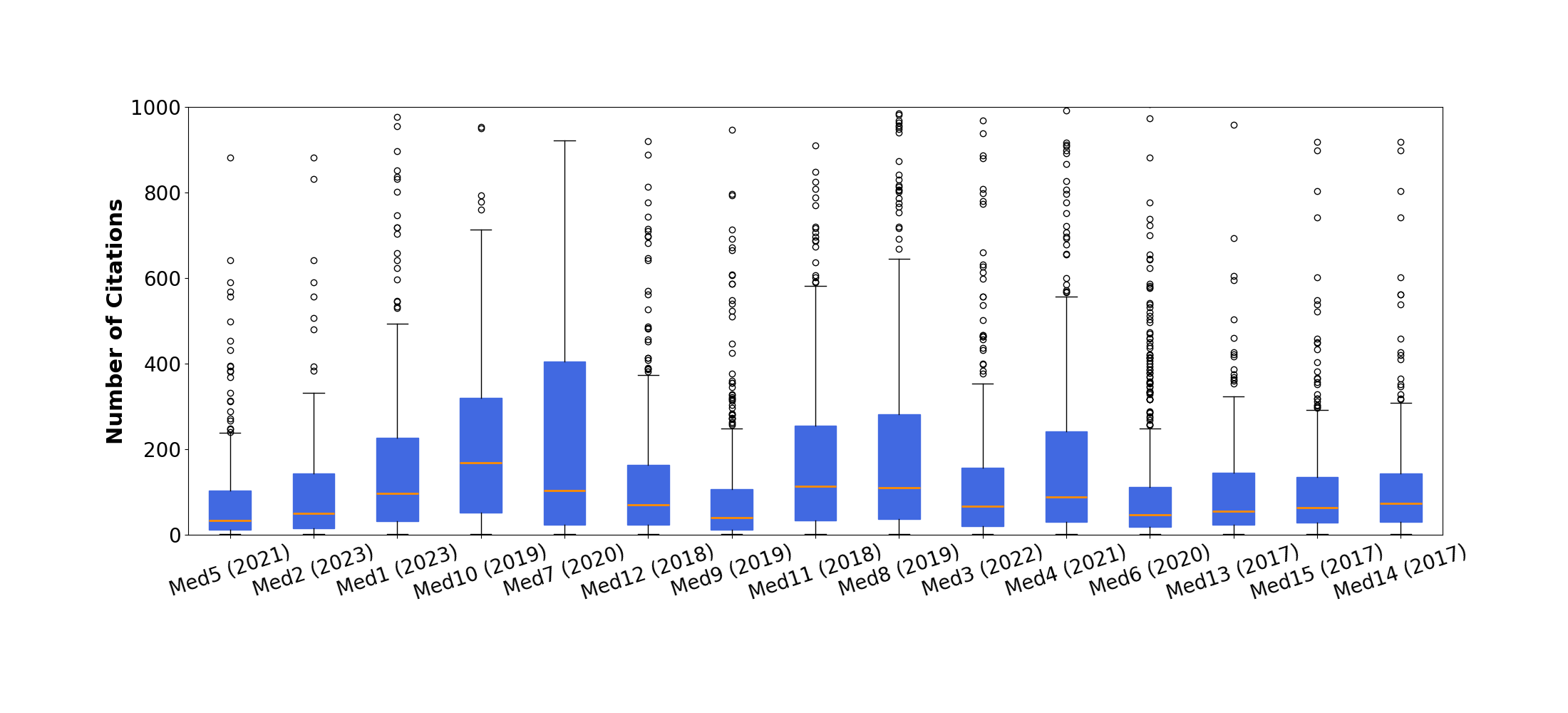}
         \caption{Box-plot of the total citation count.}
         \label{fig:15b}
     \end{subfigure}
        \caption{(a) Citation diversity (red dots) of recent Nobel laureates during (2017-2023) in physiology/medicine, along with their 95\% confidence intervals (blue vertical lines) and (b) box-plot for the citation counts of each of them.}
        \label{fig:Med_new}
\end{figure}

\begin{figure}[h]
     \centering
     \begin{subfigure}[b]{0.48\textwidth}
         \centering
         \includegraphics[width=\textwidth]{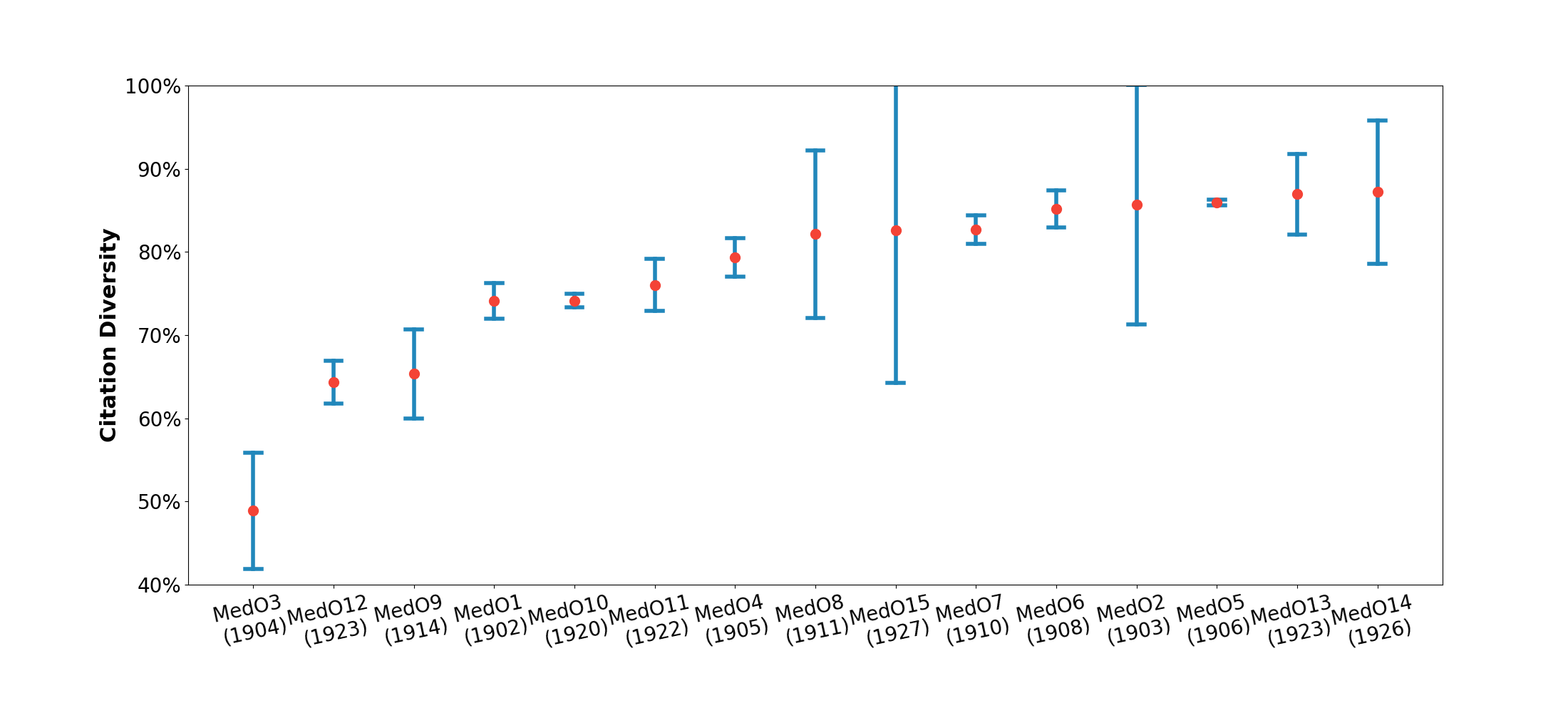}
         \caption{Citation diversity (D) with confidence intervals.}
         \label{fig:16a}
     \end{subfigure}
     \hfill
     \begin{subfigure}[b]{0.48\textwidth}
         \centering
         \includegraphics[width=\textwidth]{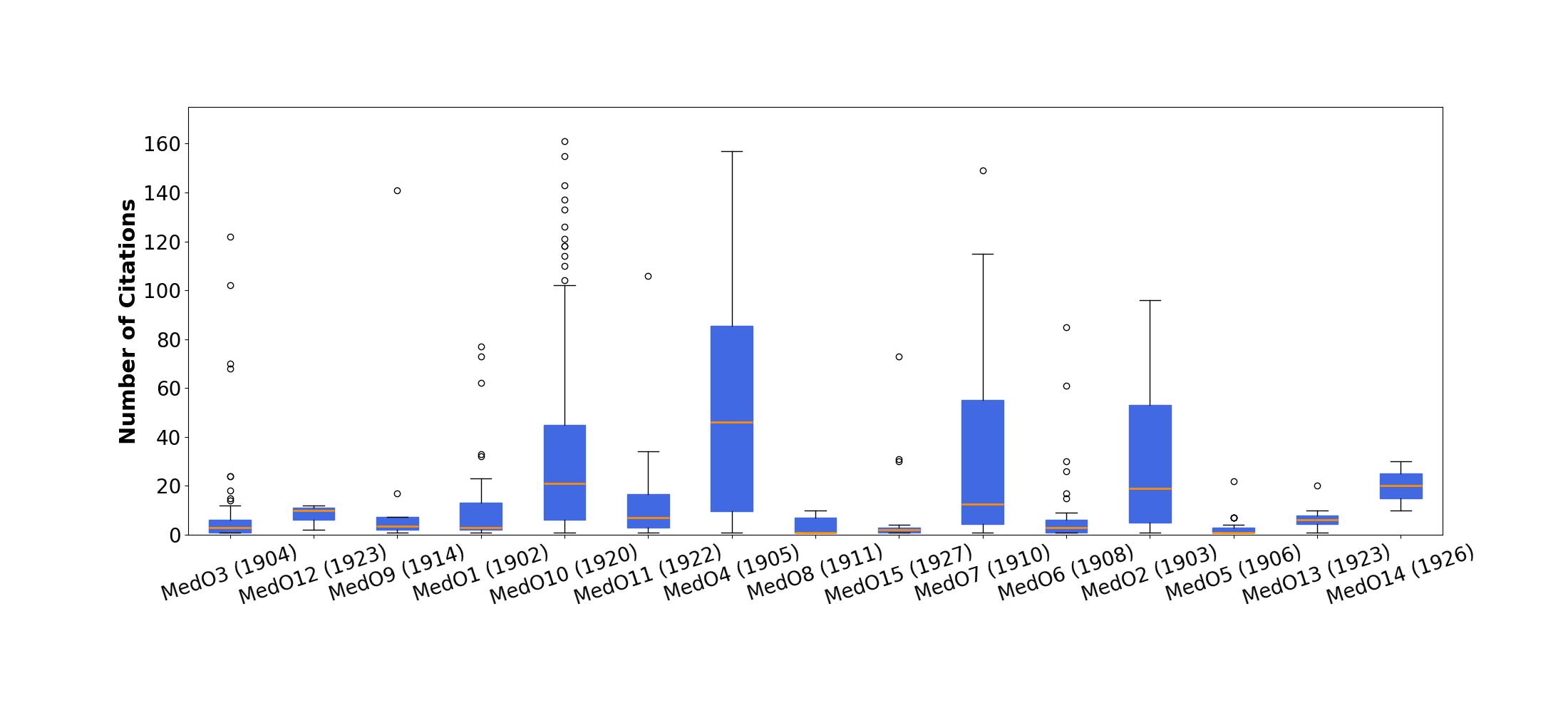}
         \caption{Box-plot of the total citation count.}
         \label{fig:16b}
     \end{subfigure}
        \caption{(a) Citation diversity (red dots) of earlier Nobel laureates during (1902-1927) in physiology/medicine, along with their 95\% confidence intervals (blue vertical lines) and (b) box-plot for the citation counts of each of them.}
        \label{fig:Med_old}
\end{figure}

In Fig.~\ref{fig:Med_new}, we show the diversity values and total citation ranges for 15 recent Nobel laureates in Physiology/Medicine. Fig.~\ref{fig:15a} categorizes them into three groups: five laureates with diversity below 85\%, eight between 85\%-90\%, and two above 90\%, indicating high citation diversity overall. The minimal confidence intervals confirm the reliability of these diversity values. 
In Fig.~\ref{fig:Med_old}, we examine diversity values and citation ranges of earlier laureates, with most diversity values falling between 65\%-87\%, except for a laureate from 1904 who shows significant low citation diversity. The broader confidence intervals for these early laureates suggest greater uncertainty in the data, unlike the more reliable and precise values seen in recent times.

In summary, this comprehensive analysis across multiple disciplines— physics, chemistry, mathematics, computer science, economics, and physiology or medicine demonstrates the significance of citation diversity values as a more insightful metric than total or average citation counts.
In physics, our analysis illustrate that diversity values have risen over time among Nobel laureates. In chemistry, despite an increase in average citation counts, diversity values have remained stable, underscoring their role in revealing citation distribution. Similar findings in mathematics reinforce the relevance of this measure.
For Turing award winners in computer science, our calculated diversity measure effectively captures citation distribution. In economics, high diversity values offering a comprehensive understanding of laureates' citation patterns. In physiology or medicine, diversity values have increased in recent times, indicating a more equitable distribution of citations among publications.
Since the awards in mathematics, computer science and economics began relatively later we cannot provide a comparative analysis, over time, of the citation diversity values in these 3 disciplines. 
Our analysis advocate the adoption of more nuanced metrics in evaluating scholarly impact, fostering a fairer assessment of academic contributions across various scientific fields.

\subsection{Citation diversity in the publication of prize winning male and female scientists}
Gender bias in paper citations plays a crucial role in making women's research less visible. Some well-documented studies \citep{Teich_2022},\citep{Sebo_2023} highlight the under-attribution of women's contributions in scientific research, evidenced by a citation gap between male and female authors. 
However, men and women still publish at similar annual rates and have comparable career-wise impact, with career length and dropout rates explaining many disparities \citep{Huang_2020}.
In an unique approach to gender-based citation analysis, our objective is to examine the uniformity in the distribution of citations of  the publications of recent award-winning male and female scientists  across six scientific disciplines using the citation diversity measure. 

\begin{figure}[h]
     \centering
     \begin{subfigure}[b]{0.4\textwidth}
         \centering
         \includegraphics[width=\textwidth]{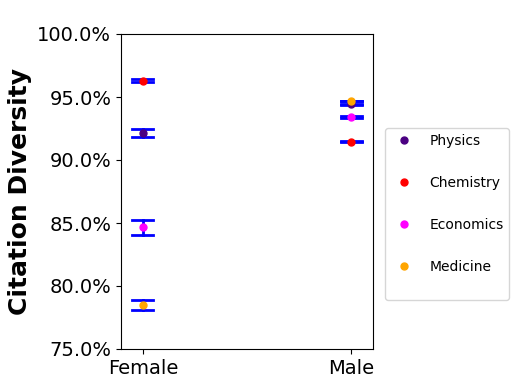}
         \caption{Citation diversity (D).}
         \label{fig:phy}
     \end{subfigure}
     \hfill
     \begin{subfigure}[b]{0.57\textwidth}
         \centering
         \includegraphics[width=\textwidth]{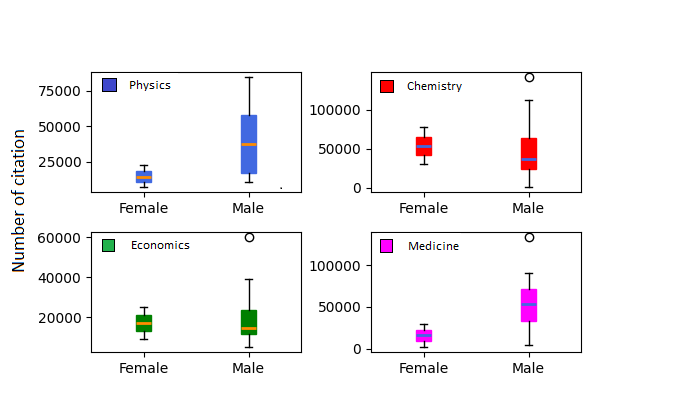}
         \caption{Box-plot of total citation count.}
         \label{fig:chem}
     \end{subfigure}
        \caption{Gender-wise citation diversity and box-plots of total citations for award winning scientists in different disciplines} 
        \label{fig:gen}
\end{figure}

Table~\ref{tab15} provides detailed information on the number of male and female award-winning scientists across different scientific disciplines, along with the period of our analysis. 
Additionally, the table includes the average citation count per publication for all male and female award winners in different  disciplines and our calculated diversity values for these scientists. It is noted that  among 126 recent award winners across six disciplines, there were no female scientists in mathematics and computer science during the period under consideration. 
In a graphical representation, Fig.~\ref{fig:phy} 
reveals that although both male and female scientists exhibit high diversity values, male scientists generally have higher diversity values than female scientists in physics, economics, and physiology/medicine (in physics and physiology/medicine the diversity values of male are very close). In chemistry, however, female award winners show a more even citation distribution than their male counterparts. Both male and female scientists in physics and chemistry have high diversity values (above 90\%). Conversely, in economics and physiology/medicine, there is a significant difference in diversity values with female scientists. Additionally, Fig.~\ref{fig:chem} depicts the total citation for male and female award winners in these four disciplines. Given the greater number of male scientists, their total citation range is higher. However, when examining the average citation count per scientist, female scientists in chemistry have a higher average, supporting the diversity value findings. While the average citation count and total citation range provide some insights, the diversity values more effectively illustrate citation distribution among male and female award winners in each discipline.

\begin{table}[h]
 \centering
 \small
\begin{tabular}{|l|c|c|c|r|r|c|c|}
\hline
Discipline & \shortstack[lb]{~\\Period of\\Analysis} & \multicolumn{2}{c|}{\shortstack[lb]{Number of \\Scientists}} & \multicolumn{2}{c|}{\shortstack[lb]{$N_{c}$\\~}} & \multicolumn{2}{c|}{\shortstack[lb]{D\\~}}\\
 \cline{3-8}
 & & Male & Female & Male & Female  & Male & Female\\
\hline
Physics&	(2017-2023)&	18	&3&	39509.11&	14823.00&	94.45&	91.93\\
\hline
Chemistry&	(2015-2023)&	17&	4&	49810.94&	54145.25&	91.42&	96.66\\
\hline
Mathematics&	(2007-2023)	&21&	0&	7117.48&-&	89.89&-\\
\hline
Computer Science&	(2010-2023)	&21&	0&	54503.90&-&74.22&-\\
\hline
Economics&	(2013-2023)&	19&	2	&19687.58&	16950.50	&93.40&	85.12\\
\hline
Physiology/Medicine&	(2014-2023)&	18	&3&	54248.83&	16066.33&	94.80&	78.28\\
\hline
\end{tabular}
\caption{Average citation count ($N_{c}$) per scientist and citation diversity (D) of male and female award winners in various 
disciplines} 
\label{tab15}
\end{table}

\section{Conclusion}
Our extensive study on citation analysis 
sheds light on various aspects of global citation diversity, offering a detailed understanding of citation patterns across different countries and academic disciplines. 
Key highlights of our work may be summarized as follows: 
\begin{itemize}
    \item 
     {\it Distribution pattern of citation counts:} We have examined the distribution of citation counts among top institutes across various countries, revealing the Pareto law nature of the upper end of the distribution with a breakdown at the lower end. It has also been  showed how the Pareto law's scaling exponent changes with the number of institutes considered across the globe. 
   \item 
     {\it Novel citation diversity measure:} A novel log-normal entropy (LNE) has been used to measure citation diversity in our analysis. Previous researches have extensively explored diversity measures across various fields, employing different metrics to assess citation distributions. However, this study marks the first instance of using an entropy-based diversity measure specifically to quantify citation distribution. We have utilized this innovative metric to effectively measure citation diversity, enhancing our understanding of the disparities in citation patterns. 
    \item 
    {\it Institutional citation diversity measure across the world:} We calculated citation diversity measures with confidence intervals, grouping countries based on these measures with respect to top few (10, 20, or 50) institutes worldwide. This revealed that many small countries share groups with large economic powers, and these groupings shift with the number of institutes considered. Box-plots have been utilized to study the total number of citations, suggesting the emergence of subgroups based on citation counts.
    \item 
    {\it Discipline-wise citation diversity:} We further calculated citation diversity along with total citation counts of award winning scientists in six disciplines (21 scientists from each discipline), physics, chemistry, mathematics, computer science, economics and physiology/medicine, uncovering the importance of measuring citation diversity of award winners across disciplines. Time evolution of the citation diversity across disciplines over the century has also been studied in three main disciplines (physics, chemistry and physiology/medicine).
    \item 
    {\it Citation diversity of publications of award winning individual scientists:} Citation diversity measures have analyzed for publications by award winning scientists in six disciplines (from 2000-2023 with publicly available data of 15 scientists from each disciplines, physics, chemistry, mathematics, computer science, economics and physiology/medicine), showing significant variation across fields. This has also been extended to individual award winners from 1901-1920 in three principal disciplines. The time evolution of author-wise diversity measures in three disciplines (physics, chemistry and physiology/medicine) provides insights into how citation patterns change over time. 
    \item 
    {\it Gender-based study in citation diversity:} Finally, a gender-based analysis of citation diversity, during the period  2007-2023, has  been done for male and female scientists in four disciplines (physics, chemistry, economics and physiology/medicine). The absence of female award winners in two disciplines (mathematics and computer science) has been noted in the considered  period of our analyses. 
\end{itemize}

This extensive study, based on the data of the top institutes or highly acclaimed elite researchers, underscores the complexity and diversity of citation practices across scientific landscapes, offering a detailed examination from multiple dimensions and perspectives. Our findings suggest that the new measure of citation diversity serves as a vital metric to assess the unevenness of the citation distribution, providing exceptional insights that citation counts alone cannot achieve.
The diversity measure, D (ranging from 0 to 100) quantifies the uniformity in the distributions of the citation patterns. Higher values of D indicate more balanced distributions, while lower values suggest concentration among a few institutions, or a few research articles  as the case may be.
As a future research project, to portray such citation  diversity analysis of the entire scientific community, one may incorporate the data from a larger and more diverse group of scientists, beyond just the elite group of top award winners. Further, our findings could be compared with established inequality indices to gain deeper insights into the structural unevenness within the scientific community.  
Additionally, investigating the lower end of the citation distributions, either in isolation or in conjunction with other distributional models that adequately fit the overall data, is another important open research question for future work.
This could provide deeper insight into the factors driving lower citation counts and shed light on the dynamics that govern citation disparities.

\section*{Acknowledgement}
We have submitted the same manuscript on arXiv as a pre-print version~\cite{arXiv}. The authors appreciate constructive suggestions of three anonymous reviewers.

\bibliographystyle{abbrv}
\bibliography{ref_revised_1}

\end{document}